\documentclass[aps,twocolumn,showpacs]{revtex4}
\usepackage{graphicx,wrapfig}
\usepackage{hyperref}
\usepackage{amsfonts}
\usepackage{amssymb}
\usepackage{amsbsy}
\usepackage{amsmath}
\usepackage{mathrsfs}
\usepackage{latexsym}
\usepackage{natbib}
\usepackage{bm}
\usepackage{subfigure} 
\usepackage{color}
\usepackage{wasysym}
\usepackage{mathbbol}
\usepackage{bigints}
\usepackage{physics}

\renewcommand*{\l}{\lambda_{\star}}

\def\o{\omega}

\def\no{\nonumber}

\def\d{\delta}

\def\p{\partial}
\def\f{\frac}

\def\th{\theta}

\def\l{\lambda}
\def\be{\begin{equation}}
\def\ee{\end{equation}}
\def\bea{\begin{eqnarray}}
\def\eea{\end{eqnarray}}

\def\p{\partial}

\newcommand{\levicivita}{}
\def\levicivita#1#{\tensor#1{\epsilon}}

\begin{document}

\title{Shadow of quantum extended Kruskal black hole and its super-radiance property}

\author{Saraswati Devi}
\email{sdevi@iitg.ac.in}

\author{Abhinove Nagarajan S}
\email{n.abhinove@iitg.ac.in}

\author{Sayan Chakrabarti}
\email{sayan.chakrabarti@iitg.ac.in}

\author{Bibhas Ranjan Majhi}
\email{bibhas.majhi@iitg.ac.in}

\affiliation{Department of Physics, Indian Institute of Technology Guwahati, 
Guwahati 781039, Assam, India}


\date{\today}

\begin{abstract}
We perform the shadow calculation of the loop quantum gravity motivated regular black hole recently proposed by Ashtekar, Olmedo and Singh (will be termed AOS black hole hereafter). In the process, we also construct the rotating loop quantum gravity inspired solution of the originally proposed static spherically symmetric AOS black hole by applying the modified Newman-Janis algorithm.  We study the quantum effects on the shadows of both the non-rotating and rotating loop quantum black hole solutions. It is observed that the general shape of the shadow for non-rotating AOS black hole is circular in shape as is expected for its classical counter part too, but the presence of loop quantum gravity inspired modification contracts the shadow radius and the effect reduces with the increase in the mass of the black hole. On a similar note, in the rotating situation, we find contraction in shadow radius due to quantum effects and the tapered nature of the shadow as expected from the classical Kerr case. However, instead of the symmetrical contraction, like non-rotating one, we found more contraction on one side relative to the other when we compare our result with the shadow of the Kerr black hole. We finally studied super-radiance in rotating AOS background and observed that the super-radiance condition for massless scalar field is identical to that of the Kerr case with the rotation of the BH being more compared to Kerr in the low mass regime. With an increase in mass of the rotating AOS black hole, the difference from Kerr starts to become insignificant.
\end{abstract}

\maketitle
\section{Introduction}\label{Introduction}
Black hole (BH) solutions are ubiquitous in the theory of gravity. In fact it is believed that every galaxy contains one supermassive BH at its centre. Several observational evidences have confirmed this, including the one in our galaxy. From the very beginning of Einstein's general theory of relativity, vast attention has been given to these compact objects, starting from various observational to theoretical aspects and their analyses. While its existence is now unquestionable, its detection, since nothing can escape from it, still remains a very tough job.  In this direction, the concept of ``black hole shadow'' becomes very important. The initial idea about the shadow of a black hole was given by Synge \cite{Synge:1966okc} and Luminet \cite{Luminet:1979nyg}, for a Schwarzschild BH and later Bardeen \cite{Bardeen} extended the idea for the Kerr BH. Recently, the event horizon telescope has observed the shadow of the BH in the centre of the $M87^*$ galaxy \cite{Akiyama:2019cqa,Akiyama:2019fyp,Akiyama:2019eap}. Till the emergence of this concept, huge attention has been devoted to find the characteristics of shadows of different BHs (for a review and more references on the development of the subject, see \cite{Cunha:2018acu,Perlick:2021aok}), for various reasons. In general it was found out that for a wide range of rotating BHs, the shadow radius is dependent on the black hole spin parameter, the configuration of the light emission region near the BH and on the angle of inclination \cite{PhysRevD.80.024042}. Studies along different avenues such as: shapes of black hole shadows with various configurations and in various background geometries \cite{1,2,3,Johannsen:2015qca,8,Perlick:2018iye, Khodadi_2022_axion}, non-rotating and rotating BH spacetime shadows in several modified theories of gravity \cite{4,5,6,10,11, https://doi.org/10.48550/arxiv.2012.05186, https://doi.org/10.48550/arxiv.2210.17533}, constraining and measuring BH parameters from the study of shadows \cite{7,Tamburini:2019vrf,Wei:2019pjf, Kumar_2020, Khodadi_2020, https://doi.org/10.48550/arxiv.2205.07787, Khodadi_2022, https://doi.org/10.48550/arxiv.2209.12584, Vagnozzi_202201}, BH shadows in dynamically evolving spacetimes \cite{9}, testing the general theory of relativity using BH shadows \cite{Vagnozzi:2019apd,PhysRevD.100.044057,Psaltis:2018xkc}, proposing new methods of calculating black hole shadows \cite{12, Chang_2020}, shadows of quantum corrected black holes \cite{13,Liu:2020ola, Khodadi_2021, https://doi.org/10.48550/arxiv.2207.02106}, proposals to use shadows as standard rulers \cite{Tsupko_2020, Vagnozzi_2020} were already done in the literature.  On another hand, it is to be mentioned that the shadow of an object in the sky does not always necessarily mean to be that of a BH, it may be of some exotic compact objects too \cite{Cardoso:2019rvt,Peng:2018nkj,Gyulchev:2019tvk,PhysRevD.100.024014,Bhattacharya:2017chr,https://doi.org/10.48550/arxiv.2208.01995} and all the above analyses can be done with such shadows which will give us important information about these objects too. 

In strong gravitational regime, the quantum nature of spacetime is very important to construct a viable theoretical model of the dynamics of gravity. Furthermore, the singularity inside the BH has been a troubling and uncomfortable region. Non-singularly complete solutions, such as regular BHs are one of the suitable candidates to avoid such situations. There exists already many regular BH solutions in the literature \cite{bardeentibilisi,Dymnikova:1992ux,Mars:1996khm,Cabo:1997rm,Bogojevic:1998ma,AyonBeato:2000zs,Hayward:2005gi}. However, in most of the cases, such BH space-times are not obtained as a solution of some underlying theory, neither are they connected to any quantum theory of gravity. On the other hand, it is well know that near the BH singularity the quantum effects are not negligible and must be incorporated within the solution itself. Towards this direction, Loop Quantum Gravity (LQG) turned out to be one of the few successful attempts to understand the quantum nature of gravity. There are few LQG inspired BH solutions \cite{Modesto:2004xx,Modesto:2005zm,Peltola:2008pa,Modesto:2009ve,Caravelli:2010ff} in literature and the characteristics of their shadows have been studied both for non-rotating as well as rotating \cite{Liu:2020ola} cases. The investigation shows the presence of quantum effects on the shadow of such quantum corrected BHs. The shadow radius contracts compared to the usual situation when one incorporates the LQG correction and  squeezes more and more as one goes to stronger quantum regime.

Very recently, Ashtekar, Olmedo and Singh \cite{Ashtekar:2018lag,Ashtekar:2018cay,Ashtekar:2020ckv} found  a complete regular static BH spacetime from an effective LQG motivated theory which is a quantum extended version of Kruskal geometry. The usual singular point $r=0$ is hidden within a minimum area element. Throughout the rest of the paper, we will denote this background as AOS BH. Not many works have been done on this particular BH background except \cite{Daghigh:2020fmw}, where the authors have studied scalar perturbations of the AOS BH and found significant difference in the quasinormal mode frequencies when compared with the Schwarzschild one. The purpose of the present paper is to understand the quantum correction through its effect on the BH shadow. In the process, we also find the rotating counter part of the static AOS BH by employing the Newman-Janis (NJ) algorithm. Then the shadows for both non-rotating and rotating BHs are obtained. We observe that the LQG inspired corrections can provide noticeable effect on the shadow only when the BH is of the order of Planck size. Therefore at present, in all practical situations, the quantum effects should remain non-detectable. Nevertheless, new physics can appear at different length scales in the theory. Having said this, completely from the theoretical aspect, we investigate the shadows at the Planck scale. On the other hand the microscopic black holes are important at primordial level. Therefore understanding these LQG inspired BHs might be relevant to understand few important aspects of the inflationary era during the early stages of our Universe.

We have the following observations.
\begin{itemize}
	\item Although the shape of the shadow for non-rotating AOS BH on the celestial plane does not change (circular in shape) when compared with the Schwarzschild BH, the presence of LQG inspired quantum modification contracts the radius. The contraction becomes less with the increase in the mass of the AOS BH.
	
   \item In rotating situation, we again find contraction in shadow radius due to quantum effects. Instead of symmetrical contraction, like non-rotating one, here we have more contraction on the right hand side relative to the left hand side when one compares with shadow for vanishing quantum parameters. 
   
   \item For a fixed mass of BH contraction increases on the right hand side while the same decreases on the left side as we increase the rotation parameter. 
   
   \item For a fixed rotation of BH the contraction decreases as the mass of BH increases. Moreover, it approaches towards shadow of the Kerr BH as one increases the mass. 
\end{itemize}

As an addition, we describe the super-radiance property of the rotating AOS BH which we have constructed as well. Black hole superradiance and stability has been studied in detail in various contexts and similar works can be found in the literature \cite{https://doi.org/10.48550/arxiv.2002.10496, Franzin_2021, Khodadi_2021s1, Khodadi_2021s2, Richarte_2022, https://doi.org/10.48550/arxiv.2201.02220, Mascher_2022, Ishii_2022, https://doi.org/10.48550/arxiv.2208.13176, Chen_2022, https://doi.org/10.48550/arxiv.2103.04239, Cuadros_Melgar_2021, Khodadi_202201}. We will show that the behaviour of the horizon angular velocity of the Kerr BHs and the rotating AOS BHs are similar and matches with each other for small spin parameter, while, the value of the angular velocity starts to change significantly as we start to increase the spin parameter.

The organisation of the paper is as follows. In the next section we briefly discuss about the non-rotating AOS BH. Using NJ formalism in this section, we also find the rotating counter part of the static AOS BH. Section \ref{Shadow} has been devoted to build the working formulae to construct the shadow. Here we show and discuss about the structures of the shadows for both non-rotating and rotating cases. The super-radiance phenomenon is discussed in Section \ref{Super}. Finally, we conclude in Section \ref{Con}. We use units where $G = c = 1$ unless otherwise specified. 

\section{AOS black hole}
AOS BH is a complete regular static solution of LQG motivated effective theory \cite{Ashtekar:2018lag,Ashtekar:2018cay,Ashtekar:2020ckv}. This does not contain the singular point $r=0$ as it has been hidden by a minimum area element, determined by ``some'' underlying microscopic theory. The original solution is spherically symmetric and non-rotating. Here we first briefly review the non-rotating AOS BH and then using the Newman-Janis (NJ) formalism its rotating counter part will be found out.

\subsection{Non-rotating AOS: a brief review}
The  effective  metric, exterior to trapping and anti-trapping horizons, is given by static, spherically symmetric form \cite{Ashtekar:2018lag,Ashtekar:2018cay,Ashtekar:2020ckv}:
\begin{equation} 
g_{ab} d x^{a} d x^{b} =  - \f{p_b^2}{p_c\, L_o^2} d x^2 + \f{\gamma^{2} p_{c}\, \delta_{b}^{2}}{\sinh^{2} (\delta_{b}b)} d T^2+ p_c d \omega^{2}~,
\label{qm1}
\end{equation}
where $x$ and $T$ are the time and radial coordinates, respectively and $d\omega^2$ is the metric on a unit $2$-sphere.
The parameters appearing in metric coefficients are determined as
\begin{align}
&\tan\Big(\frac{\delta_c{c}\left(T\right)}{2}\Big)=\frac{\gamma L_0\delta_c}{8m}e^{-2T}~,
\no
\\
&{p}_c\left(T\right)=4m^2\Big(e^{2T}+\frac{\gamma^2L_0^2\delta_c^2}{64m^2}e^{-2T}\Big)~,
\label{cpc}
\no
\\
&\cosh \big(\delta_{b }\,b(T)\big) = b_o \tanh\left(\f{1}{2}\Big(b_o T + 2 \tanh^{-1}\big(\frac{1}{b_o}\big)\Big)\right), \no\\
&p_b(T) = -2m\gamma L_0\,  \f{\sinh \left(\delta_{b}\, b(T)\right)}{\delta_{b}}\, \f{1}{\gamma^{2 }-\f{\sinh^2\left(\delta_{b}\, b(T) \right)}{\delta_{b}^2} }~,
\end{align}
where $m = \f{GM}{c^2} = M$ (in the units mentioned above) is the mass parameter and
$b_{o}^{2} = 1 + \gamma^{2}\delta_{b}^{2}$. Here $\delta_{b}$ and  $ \delta_{c}$ are the quantum parameters given by,
\begin{equation}
\d_{b}=\Big(\frac{\sqrt{\Delta}}{\sqrt{2\pi}\gamma^2m}\Big)^{1/3}~; \,\,\,\ 
L_{o}\d_{c}=\frac{1}{2} \Big(\frac{\gamma\Delta^2}{4\pi^2 m}\Big)^{1/3}~.
\label{qpp}
\end{equation}
In the above,
$\Delta$ is the minimum non-zero eigenvalue of the area operator in LQG, given by  $\Delta \approx 5.17 \ell_{pl}^{2}$ and
$\gamma \approx 0.2375$ is the Barbero-Immirzi parameter.
$L_{o}$ is an infrared regulator introduced to make the phase space description well-defined. The location of horizon is determined by $T=0$.

In order to express the above one in our familiar static, spherically symmetric form 
\begin{equation}
ds^2=-f(r) dt^2+\frac{dr^2}{g(r)}+h(r)\left(d\theta^2+\sin^2\theta d\phi^2\right)~, 
\label{sss1}
\end{equation}
in Schwarzschild like coordinates the following change of notations has been considered in \cite{Ashtekar:2020ckv}:
\begin{equation}
 t = x,\quad  r_{S} = 2m,\quad  r= r_{S}\,e^T,  \quad b_{0} \equiv (1 +\gamma^{2}\delta_{b}^{2})^{\f{1}{2}}
= 1 + \epsilon~.
\end{equation}
In this case the metric coefficients are identified as \cite{Ashtekar:2020ckv}
\begin{eqnarray}
-f(r)&=& -\left(\frac{r}{r_S}\right)^{2\epsilon}\frac{\left(1-\left(\frac{r_S}{r}\right)^{1+\epsilon}\right)\left(2+\epsilon+\epsilon\left(\frac{r_S}{r}\right)^{1+\epsilon}\right)^{2} }{16\left(1+\frac{\delta_{{c}}^{2} L_0^{2}\gamma^{2}r_S^2}{16  r^4} \right) (1+\epsilon)^{4}}
\nonumber
\\
&\times& \left((2+\epsilon)^{2}-\epsilon^{2}\left(\frac{r_S}{r}\right)^{1+\epsilon}\right)~;
\label{fr}
\end{eqnarray}
\begin{eqnarray} 
\f{1}{g(r)} &=& \Big(1+\frac{\delta_{{c}}^{2} L_0^{2}\gamma^{2}r_S^2}{16 r^4} \Big)
\nonumber
\\
&\times& \frac{\Big(\epsilon +\left(\frac{r}{r_S}\right)^{1+\epsilon } (2+\epsilon )\Big)^2}{\Big(\left(\frac{r}{r_S}\right)^{1+\epsilon }-1\Big) \Big(\left(\frac{r}{r_S}\right)^{1+\epsilon } (2+\epsilon )^2-\epsilon ^2\Big)}~;
\label{gr}
\end{eqnarray}
and
\begin{equation}
h(r) = 4m^2\left( e^{2T}+\frac{\gamma^2L_0^2\delta_{c}^2}{64m^2}e^{-2T}\right) = r^2\left(1+\frac{\gamma^{2}L_{0}^{2}\delta_{c}^{2} r_S^2}{16  r^4}  \right)~.
\label{hr}
\end{equation}
Henceforth the quantum parameters are $\delta_{c}$ and $\epsilon$ (since we switched from $\delta_{b}$ to $\epsilon$). This form of the metric will be suitable for our main purpose.

Few important features of the form (\ref{sss1}) are worth mentioning here. $f(r)$ diverges as one goes to $r\to\infty$. But still AOS can be shown to be asymptotically Minkowski. As mentioned in \cite{Ashtekar:2020ckv} while one generally checks asymptotic flatness by checking the $r \to \infty$ limit of the metric components, and comparing with $\eta_{ab}$, one can show that a given metric $g_{ab}$ can be taken to be asymptotically flat at spatial infinity if the components reduce at least as fast as $1/r$ as $r \to \infty$, to the components of some flat metric $\tilde{\eta}_{ab}$ which exists, while the coordinates $(t,\theta, \phi)$ are kept constant. In \cite{Ashtekar:2020ckv}, they show asymptotic flatness of the above metric using this idea and a time dependent coordinate change. Then, this time coordinate is no longer associated with a timelike Killing vector. For the detailed analysis see Sec IV.B. of \cite{Ashtekar:2020ckv}. 

\subsection{Rotating spacetime through modified Newman-Janis Algorithm}
The usual Newman-Janis algorithm (NJA), as proposed in  \cite{Newman:1965tw,Newman:1965my}, is used to  construct  a stationary and axisymmetric spacetime from a static and spherically symmetric one having the form (\ref{sss1}). The steps involved in the algorithm have been discussed in Appendix \ref{Appendix A}. This formalism has been successfully used in several cases to find the rotating counter part of the static spherically symmetric metric.
So, applying NJA to the metric (\ref{sss1}), one can in principle obtain the rotating counterpart of the same. However the job of 
choosing the exact complexified form  of the metric functions is a tedious one because there are various ways in which this
can be done and it also needs to be chosen 
in such a way that the transformation (\ref{GT}) is allowed i.e., $\chi_{1}(r)$ and $\chi_{2}(r)$ must be functions of $r$ only and not any other coordinates. The usual procedure fails to satisfy this for our present metric (\ref{sss1}) (for details see Appendix \ref{AppA1}).

To do away with this, we resort to the modified version of NJA as was proposed by Azreg-A\"{\i}nou's non-complexification procedure \cite{PhysRevD.90.064041}, where the modification is incorporated in the third step (rest all steps are same). As per this new procedure, the complexification of the radial coordinate $r$ is simply dropped and instead of that we consider $\d_{\nu}^{\mu}$, in Eq. (\ref{TV}), transform as a vector under the transformation (\ref{com}). In that case the metric coefficients $f(r), g(r)$ and $h(r)$ transform to $F=F(r,a,\theta)$, $G=G(r,a,\theta)$ and $H=H(r,a,\theta)$, respectively. The final form of the rotating metric in Boyer Lindquist coordinates after applying the modified NJA becomes
(see Appendix \ref{AppA2}):
\begin{eqnarray}
&&ds^{2}=-Fdt^{2}-2a\sin^{2}\theta\bigg(\sqrt{\f{F}{G}}-F\bigg)dt d\phi+\f{H}{\Delta(r) }dr^{2}
\nonumber
\\
&+&H d\theta^{2}+\sin^{2}\theta \bigg[H+a^{2}\sin^{2}\theta\bigg(2\sqrt{\f{F}{G}}-F\bigg)\bigg]d\phi^{2}~,
\label{BLC}
\end{eqnarray}
where 
\begin{equation}
\Delta(r) = GH+a^2\sin^2\theta = g(r)h(r)+a^2~;
\label{BB1}
\end{equation}
and $F, G$ are given by (\ref{F}) and (\ref{G}), respectively while $H$ remains undetermined. 
The above one represents the rotating AOS BH. We will work with this form of the metric.

\section{Finding the shadows}\label{Shadow}
\subsection{Working formulas}
We shall now discuss the equations involved in obtaining the contour of the shadow. Since shadow contours correspond to unstable circular
null geodesics, it is necessary to obtain such equations first for the metric of our case. To find the null geodesics around the AOS BH
we shall use the Hamilton-Jacobi (H-J) equation. We give here the required expression for rotating case, given by metric (\ref{BLC}). The non-rotating results can be obtained just by setting $a=0$. After some calculations (see  Appendix \ref{Appendix B}), the 
separated geodesic equations for the photon are found to be
\begin{eqnarray}
\frac{F}{G}\Delta(r)\frac{dt}{d\lambda}&=&\left[H+a^2\sin^2\theta\left(2\sqrt{\frac{F}{G}}-F\right)\right]E
\nonumber
\\
&-& a\left(\sqrt{\frac{F}{G}}-F\right)L~;
\label{eq:t_eqn}
\end{eqnarray}
\begin{equation}
\frac{F}{G}\Delta(r)\sin^2\theta\frac{d\phi}{d\lambda}=a\sin^2\theta\left(\sqrt{\frac{F}{G}}-F\right)E+FL~;
\label{eq:phi_eqn}
\end{equation}
and
\begin{equation}
H\frac{dr}{d\lambda}=\pm \sqrt{R(r)}~;
\label{eq:r_eqn}
\end{equation}
\begin{equation}
H\frac{d\theta}{d\lambda}=\pm \sqrt{\Theta(\theta)}~,
\label{eq:theta_eqn}
\end{equation}
where
\begin{equation}
R(r)=\left[\Sigma(r)E-aL\right]^2-\Delta(r)\left[\mathcal{Q}+\left(L-aE\right)^2\right]~,
\end{equation}
and
\begin{equation}
\Theta(\theta)=\mathcal{Q}+a^2E^2\cos^2\theta-L^2\cot^2\theta~.
\end{equation}
In the above $\Sigma(r)$ symbolizes
\begin{equation}
\Sigma(r) = \sqrt{\frac{G}{F}}H+a^2\sin^2\theta = \sqrt{\frac{g(r)}{f(r)}}h(r) + a^2~.
\end{equation}
Here $R(r)$ and $\Theta(\theta)$ must be non-negative; i.e., for the photon motion, we must have
\begin{equation}
\frac{R(r)}{E^2}=\left[\Sigma(r)-a\xi\right]^2-\Delta(r)\left[\eta+\left(\xi-a\right)^2\right]\geq 0~,
\label{eq:R}
\end{equation}
and
\begin{equation}
\frac{\Theta(\theta)}{E^2}=\eta+(\xi-a)^2-\left(\frac{\xi}{\sin\theta}-a\sin\theta\right)^2\geq 0~,
\label{eq:Theta}
\end{equation}
 In the above $\xi[ =L/E]$ and $\eta[=\mathcal{Q}/E^2]$ are the critical impact parameters (also known as Chandrasekhar's constants) that determine the motion of the photon.

As already stated, the contour of a shadow depends on the unstable light rings.
In the general rotating spacetime, these unstable circular photon orbits must satisfy ,
$R(r_{ph})=0$, $R'(r_{ph})=0$ and $R''(r_{ph}) \geq 0$, where $r=r_{ph}$ is the radius of the unstable photon orbit. From the first two conditions we have
\begin{equation}
\left[\Sigma(r_{ph})-a\xi\right]^2-\Delta(r_{ph})\left[\eta+\left(\xi-a\right)^2\right]=0~,
\label{eq:Req0}
\end{equation}
and
\begin{equation}
2\Sigma'(r_{ph})\left[\Sigma(r_{ph})-a\xi\right]-\Delta'(r_{ph})\left[\eta+\left(\xi-a\right)^2\right]=0~.
\label{eq:Rpeq0}
\end{equation}
The valid solution for $\xi$ for describing a BH shadow that is found from above is
\begin{equation}
\xi=\frac{\Sigma(r_{ph})\Delta'(r_{ph})-2\Delta(r_{ph})\Sigma'(r_{ph})}{a\Delta'(r_{ph})}~.
\label{xi}
\end{equation}
Using this and solving for $\eta$ we have,
\begin{equation}
\eta=\frac{4a^2\Sigma'^2_{ph}\Delta_{ph}-\left[\left(\Sigma_{ph}-a^2\right)\Delta'_{ph}-2\Sigma'_{ph}\Delta_{ph} \right]^2}{a^2\Delta'^2_{ph}}~.
\label{eta}
\end{equation}
where the subscript ``$ph$" indicates that the quantities are evaluated at $r=r_{ph}$. The general expressions for the critical impact parameters $\xi$ and $\eta$ of
the unstable photon orbits are given by equations
(\ref{xi}) and  (\ref{eta}). In order to obtain the apparent shape of a shadow, the celestial coordinates $\alpha$ and $\beta$ which lie in the celestial plane perpendicular to the line joining the observer and the center of the spacetime geometry are used. If the observer is situated at $(r_0,\theta_0)$, then the celestial coordinates are defined as \cite{Book1}
\begin{equation}
\alpha= -r_0^2\sin\theta_0\frac{d\phi}{dr}\Big\vert_{(r_0,\theta_0)}~,
\end{equation}
and
\begin{equation}
\beta= r_0^2\frac{d\theta}{dr}\Big\vert_{(r_0,\theta_0)}~,
\end{equation}
If the general metric is asymptotically flat, then the above equations reduce to 
\begin{equation}
\alpha=-\frac{\xi}{\sin\theta_0}~,
\label{eq:alpha}
\end{equation}
and
\begin{equation}
\beta=\pm \sqrt{\eta+a^2\cos^2\theta_0-\xi^2\cot^2\theta_0}~.
\label{eq:beta}
\end{equation}

Using Eqs. (\ref{xi}), (\ref{eta}), (\ref{eq:alpha}) and (\ref{eq:beta}),  parametric plots of $\alpha$ and $\beta$ are obtained by using the unstable photon orbit radius $r_{ph}$ as a parameter that define the contour of the shadow.
It is to be mentioned here that the expressions for $\Sigma (r)$ and $\Delta(r)$ contain the spherically symmetric static metric functions $f(r)$, $g(r)$ and $h(r)$. So to obtain the shadow for rotating metric, the information from the non-rotating spacetime will be used.
\subsection{Shadow for non rotating case}



In this section we shall discuss how the shadow contour looks for the case of a non-rotating AOS BH
i.e., for $a=0$ as described by the metric (\ref{sss1}). Also since the underlying metric now is spherically 
symmetric, so the shadow to an observer seems the same whatever be the 
value of $\theta_{0}$. Hence we can take up the simple case of $\theta_{0}=\f{\pi}{2}$.
Equation (\ref{eq:alpha}) and (\ref{eq:beta}) then reduces to
\begin{equation}
\alpha=-\xi~;\,\,\,\
\beta=\pm \sqrt{\eta}~,
\end{equation}
and so we have 
\begin{equation}
\alpha^{2}+\beta^{2}=\xi^{2}+\eta~. 
\end{equation}
With the spin parameter $a$ set to zero we have $\Delta(r)=G(r)H(r)=g(r)h(r)$ and $\Sigma(r)=k(r)=\sqrt{\f{g(r)}{f(r)}}h(r)=\sqrt{\f{G(r)}{F(r)}}H(r)$ and, from the conditions
$R(r_{ph})=0$ and $R'(r_{ph})=0$ we obtain the following equations
\begin{equation}
\eta + \xi^{2}=\f{h(r_{ps})}{f(r_{ps})}~,
\label{con1}
\end{equation}
and
\begin{equation}
f'(r_{ps})h(r_{ps})-f(r_{ps})h'(r_{ps})=0~.
\label{con2}
\end{equation}
Solving equation (\ref{con2}) we get the value of radius of the photon sphere $r_{ps}$.
Putting this value in (\ref{con1}) gives the value of $R_{sh}$, the radius
of the shadow contour:
\begin{equation}
R_{sh}=\sqrt{\alpha^{2}+\beta^{2}}=\sqrt{\xi^{2}+\eta}=\sqrt{\f{h(r_{ps})}{f(r_{ps})}}~.
\end{equation}

The shadows for the non-rotating AOS BH along with the Schwarzschild one is plotted in Fig. \ref{nr} for different values of $m$ and the quantum parameters  $\delta_{c}$  and $\epsilon$ as mentioned in Table \ref{tab1}. It is immediately seen that the shadow radius for the quantum corrected AOS BHs are always small for a whole range of mass, starting from $1 \ell_{pl}$ to $10\ell_{pl}$, compared to the shadow radius of the Schwarzschild BH. As the mass of the AOS BH increases from $1\ell_{pl}$ to $10 \ell_{pl}$, the radii also increase accordingly.   

\begin{widetext}

\begin{table}[!htbp]
 	\centering
 	\begin{tabular}{|c|c|c|c|c|c|c|}
 		\hline
 		$m$&$\gamma$&$\delta_{c}$&$\delta_{b}$&$\epsilon$&$r_{ps}$&$R_{sh}$\\
 		\hline
 		$1 \ell_{pl}$& -&0&0&0&4.84$\times10^{-35}$& 8.39$\times$$10^{-35}$ \\
 		\hline
 		$1\ell_{pl}$&0.2375&2.92$\times$ $10^{-36}$&2.52&0.16&4.99$\times$$10^{-35}$&8.17$\times$$10^{-35}$\\
 		\hline
 		$2\ell_{pl}$&0.2375&2.32$\times$ $10^{-36}$&2.01&0.11&9.87$\times$$10^{-35}$&1.6$\times$$10^{-34}$\\
 		\hline
 		$10\ell_{pl}$&0.2375&1.35$\times$ $10^{-36}$&1.17&0.03&4.87$\times$$10^{-34}$&8.34$\times$$10^{-34}$\\
 		\hline
 	\end{tabular}
 	\caption{ The table shows the values of different parameters for the non-rotating case. The first row contains the numbers for Schwarzschild black hole with mass $m=1\ell_{pl}$ and the corresponding rows contain the values for the AOS black hole with masses $1\ell_{pl},~2\ell_{pl}$ and $10\ell_{pl}$ respectively.}
 	\label{tab1}
\end{table}
\end{widetext}

\begin{figure}[h]
	\centering
	\includegraphics[width=8cm]{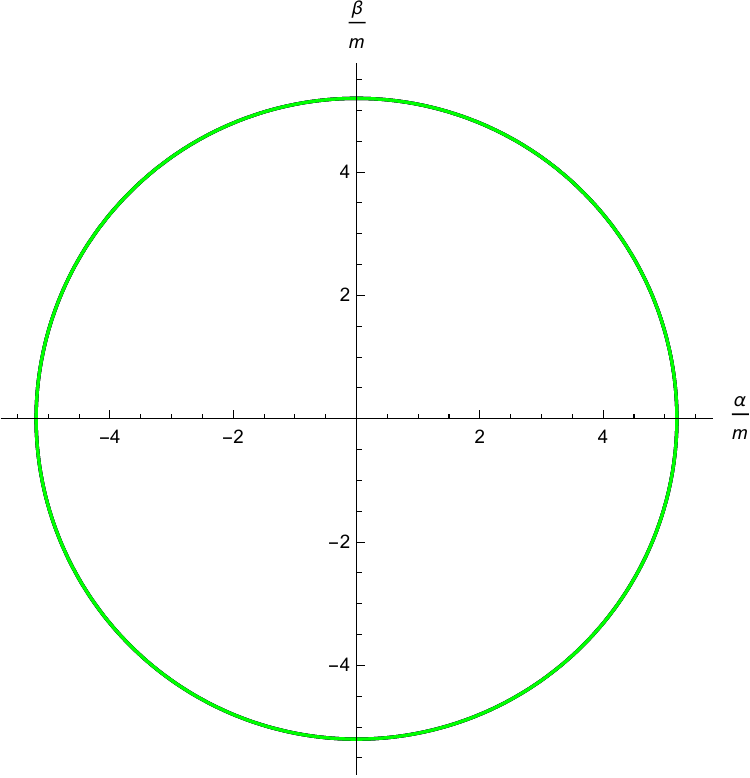} 
	\qquad
	\caption{Shadows for the classical Schwarzchild BHs for different values mass. Note that all the plots coincide since the ratios $\f{\alpha}{m},\f{\beta}{m}$ scale uniformly, independent of the mass. As in Fig \ref{nr}, the red, blue and black lines correspond to masses $m = 1 l_{Pl}, m = 2 l_{Pl}$ and $m = 10 l_{Pl}$ }%
	\label{sch}
\end{figure}

\begin{figure}[ht]
	\centering
	\includegraphics[width=9cm]{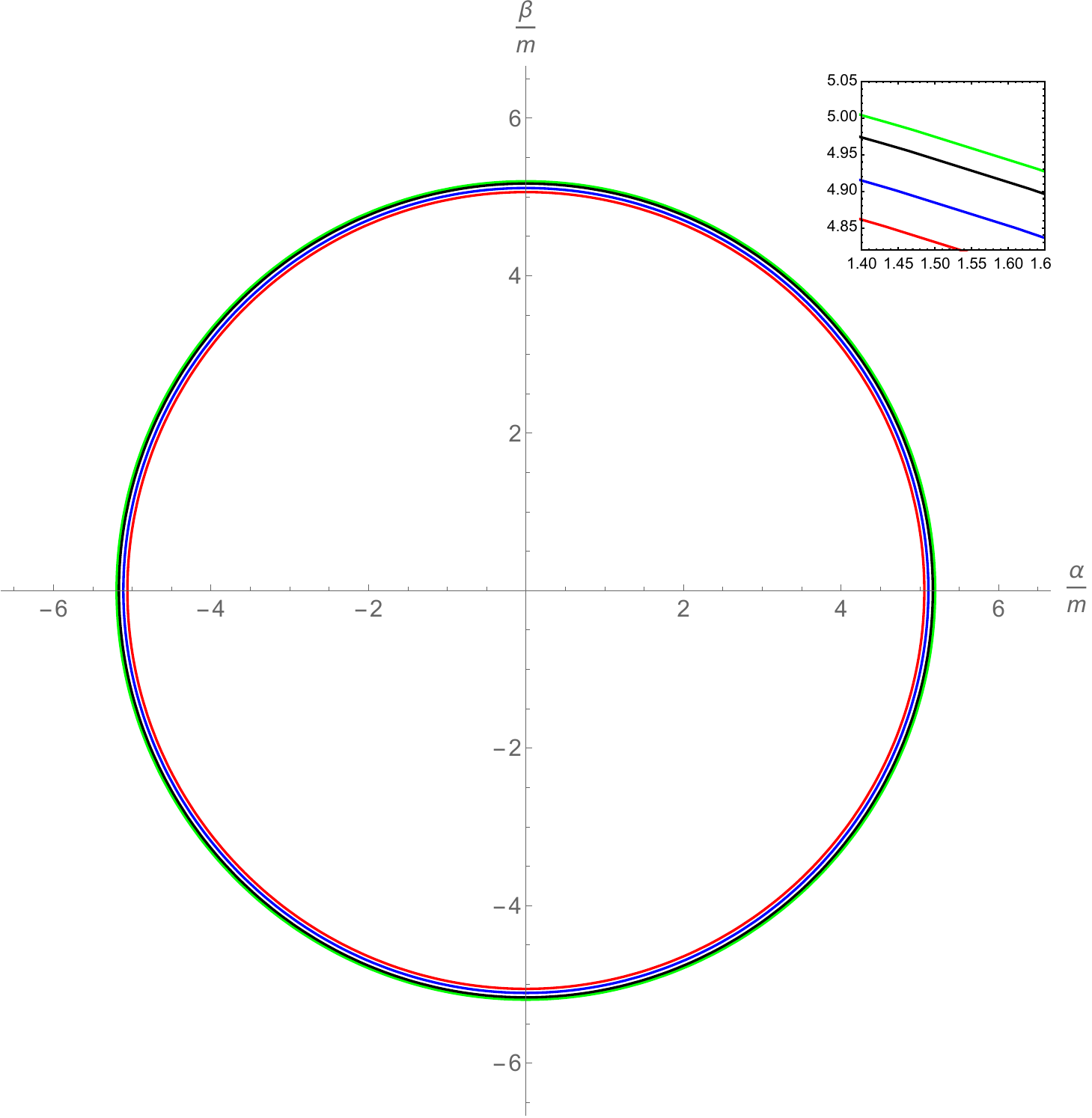} 
	\qquad
	\caption{Shadows for the non rotating AOS BH  for different values of the quantum parameters
		$\d_{c}$ and $\epsilon$. The green circle corresponds to the standard Schwarzschild BH shadow with $m=1\ell_{pl}$. The red, blue and black circles correspond to the quantum case with masses $m=1\ell_{pl}$, $m=2\ell_{pl}$ and $m=10\ell_{pl}$ respectively. A partly zoomed plot of the shadow contour in the first quadrant is plotted in the inset for understanding the difference between the standard Schwarzschild case with the AOS one.}%
	\label{nr}
\end{figure}


\subsection{Shadow for the rotating AOS BH}

We shall now apply the above formulae to obtain the contour of the shadow of a rotating AOS BH whose corresponding non-rotating counterpart is represented by the  metric (\ref{sss1}) with the metric coefficients given by (\ref{fr}), (\ref{gr}) and (\ref{hr}).
Also we want to compare and see how the shadow contour for quantum case varies with that of Kerr case. The Kerr results can be obtained by setting the quantum parameters $\delta_{c}$ and $\epsilon$ in (\ref{fr}), (\ref{gr}) and (\ref{hr}) to zero. 
\begin{figure*}[!htbp]
	\centering
	\includegraphics[width=7.5cm]{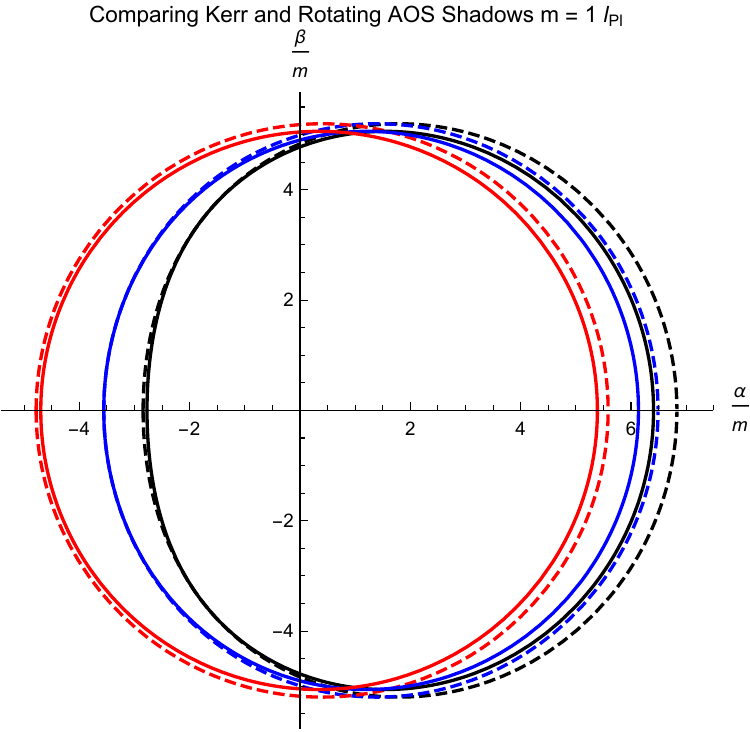}
	\qquad
	\includegraphics[width=7.5cm]{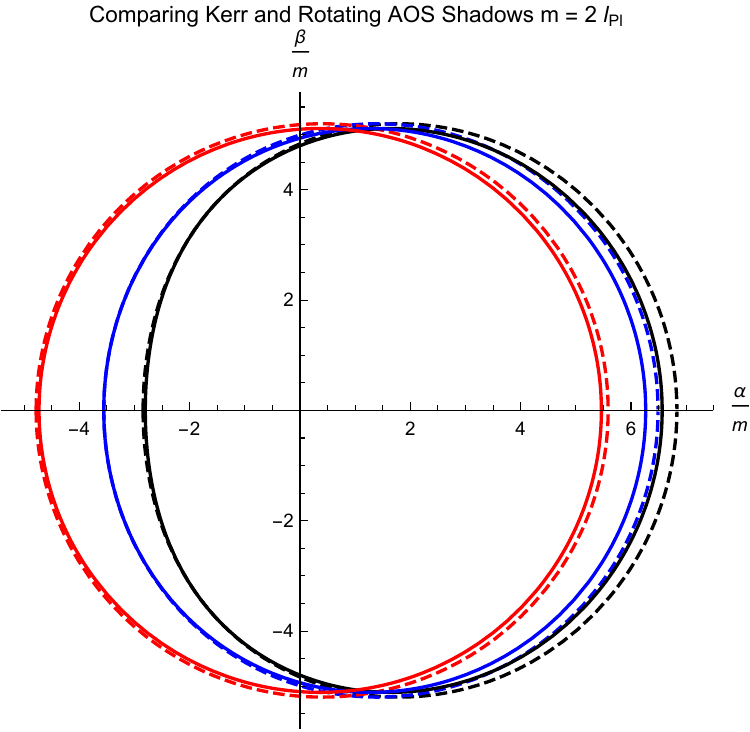}
	\qquad
	\includegraphics[width=7.5cm]{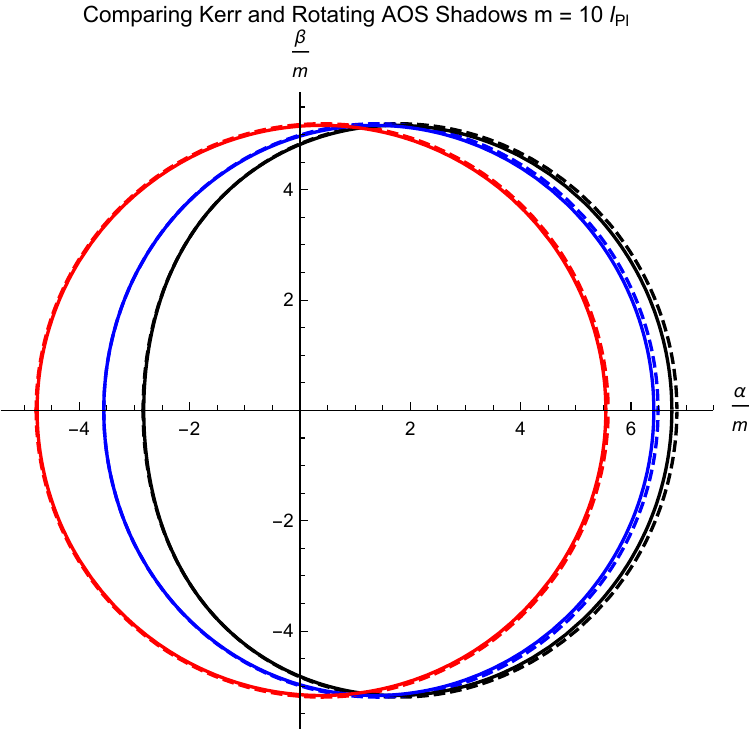}
	\caption{The figure shows the contours of shadow for different values spin
		parameter $a$ with $m=1 \ell_{pl}, 2 \ell_{pl},10 \ell_{pl}$. Color codes: Red $(a=0.2 m)$, Blue $(a=0.7 m)$ and Black $(a=0.9 m)$. The dashed contours represent the Kerr case while the solid contours represent the rotating AOS BH.}
		\label{fig1}
	\end{figure*}

In figures from Fig \ref{fig1} to Fig \ref{newfig1}, the contours of the shadow are 
shown for the rotating AOS and Kerr BHs for different values of $m$ and spin parameter $a$. 

\begin{figure}[!htbp]
	\centering
	\includegraphics[width=8cm]{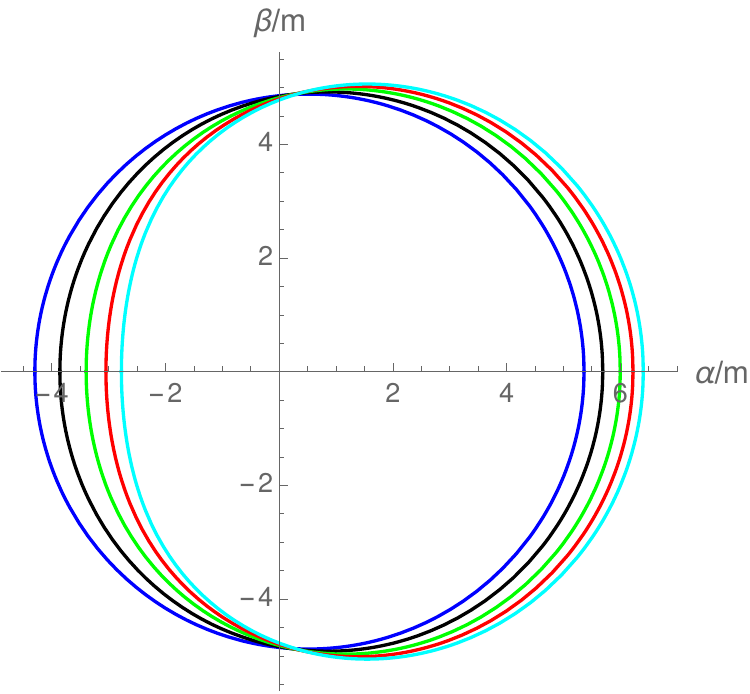}
	\caption{Shadows for the effective quantum metric with various values of inclination angles
		$\theta_{0}$ for $a=0.9 m$ with $m=1\ell_{pl}$. Color codes:  (Blue $\theta_{0}=17^{0}$), (Black $\theta_{0}=30^{0}$),
		(Green$ \theta_{0}=45^{0}$), (Red $\theta_{0}=60^{0}$), (Cyan $\theta_{0}=90^{0}$).}
	\label{fig:inc}
\end{figure}

Also the shape of shadows at different inclination angles for spin parameter $a=0.9m$ ( $m=1\ell_{pl}$ ) has been shown in Fig. \ref{fig:inc} to see the variation in shadow with varying inclination angles.

It is observed after comparing the Quantum and Kerr case in Fig. \ref{fig1} 
that the presence of quantum parameters
tends to shrink the shadow i.e., the size of the shadow decreases due to quantum effect. Note that in Fig \ref{fig1} 
the dashed contours represent the Kerr case while the solid contours represent the rotating AOS BH. As the quantum parameters $\d_{c}$ and $\epsilon$ 
are increased(i.e. for the smaller masses), both the photon sphere radius and shadow radius decreases. In Fig \ref{fig1} - Fig \ref{newfig1}, we are plotting $\f{\alpha}{m}$ and $\f{\beta}{m}$. In the Kerr and Schwarzchild cases, this quantity scales uniformly, and hence we do not see a difference with changing mass. However, in the AOS case, since the masses are a function of the quantum parameters, the scaling of these ratios are not uniform, which lead to the corrections observed in the plots. Observing these figures,it can be inferred that for a fixed mass, the shadow shrinks more on the right hand side than on the left hand due to quantum effect when compared with the  Kerr case and as the mass is increased, there is not much variation in the shadows of Kerr case and quantum AOS case.This happens because the quantum parameters $\d_{c}$ and $\epsilon$ are inversely proportional to $m$ as can be seen from equation (\ref{qpp}), so their values decrease when mass is increased and hence their impact on the shadow also reduces. With increasing values of $a$, not only the shadow size decreases but also the shape gets distorted. As can be seen, this distortion is more on the left part of the plot than on the right. While classically as well the asymmetry can be explained by frame dragging, the quantum corrections affect the rotating AOS black hole differently compared to a simple mass scaling. For a fixed value of the rotation parameter $a$, if we compare the change in mass of Kerr to that of rotating AOS as in Fig \ref{newfig1}, we find that there is a non-trivial difference in the shadow curvature at the left and right points in AOS while in Kerr, the shadow contours are unaffected. As noted earlier, we see that the deviation from the mass is more on the right side. This can be explained by looking at how the quantum corrections affect the curvature of the shadow at the left and right points. The two sides in the shadow contour generally reflect the prograde and retrograde orbits around the black hole. We can compare the curvature at the two extreme points for the Kerr and AOS case, as is done in \ref{curvfig}. Further we wish to point out that the left curvature for the AOS rotating black hole, falls faster than that of Kerr for larger $\f{a}{M}$. In Fig \ref{curvfig}, the curvature on the left crossing point crosses the X axis before $a = M$, indicated by the dashed black line. This has to do with the horizon structure of the rotating AOS black holes in this regime, and will be studied in detail in Sec IV.B.1. Also from Fig. \ref{fig:inc}, it is seen that  for a fixed value of $m$ and $a$, if the inclination angle is increased, then the deviation from circularity increases and the shadow size also becomes smaller.

\begin{figure*}[!htbp]
	\centering
	\includegraphics[width=7.5cm]{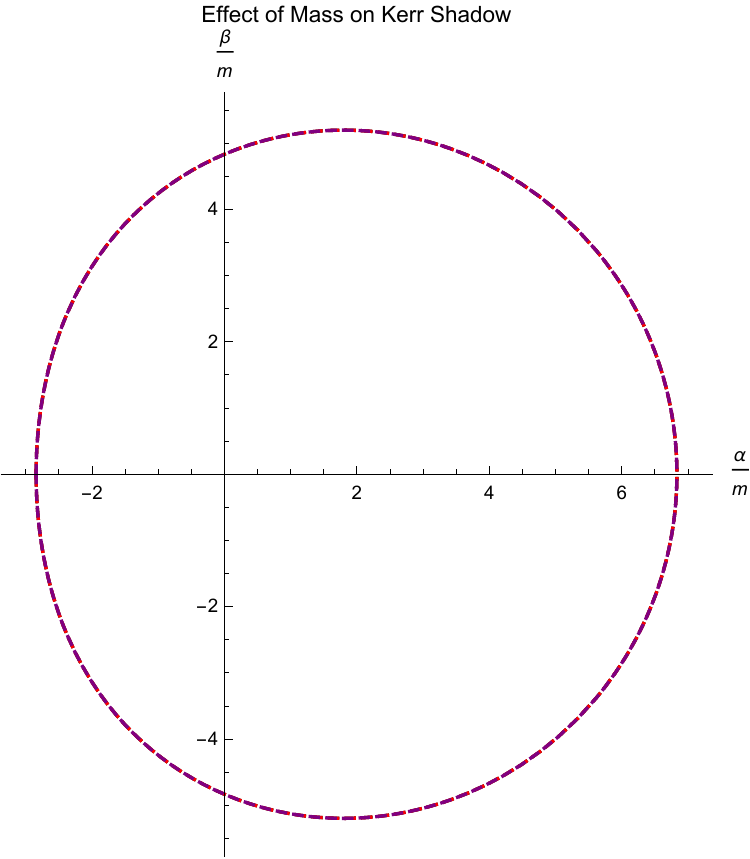}
	\qquad
	\includegraphics[width=7.5cm]{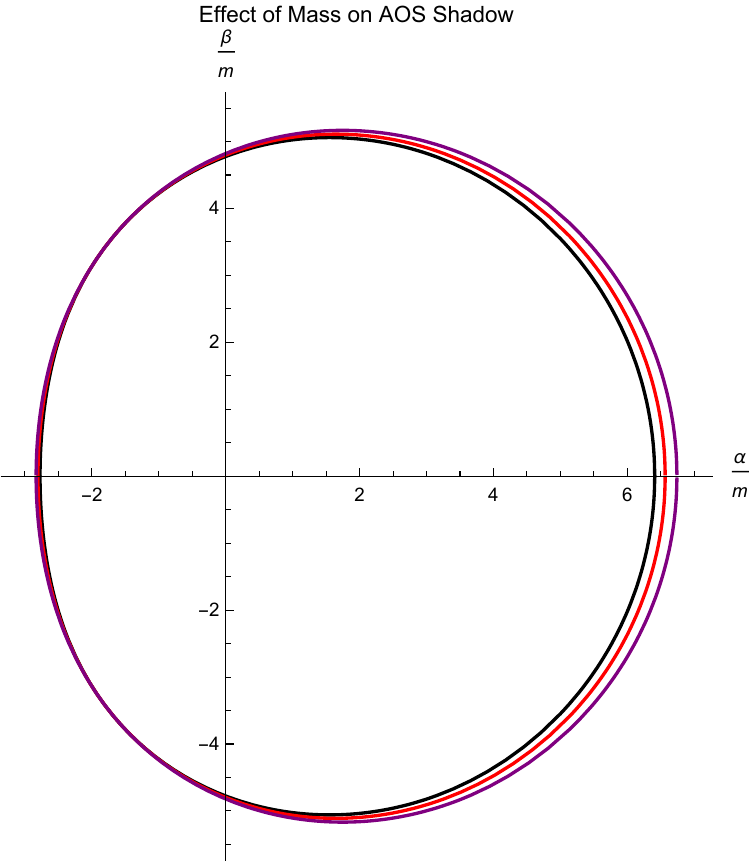}
	\caption{The figure shows the contours of shadow for a fixed spin
		parameter $a = 0.9 m$ with varying masses. Color codes: Black $(m = 1 \ell_{pl})$, Red $(m = 2 \ell_{pl})$ and Purple $(m = 10 \ell_{pl})$.}
		\label{newfig1}
\end{figure*}

\begin{figure*}[!htbp]
	\centering
	\includegraphics[width=7.5cm]{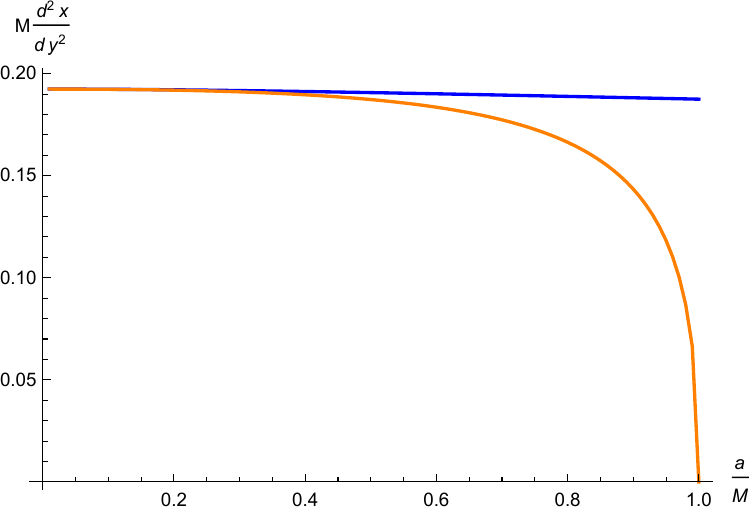}
	\qquad
	\includegraphics[width=7.5cm]{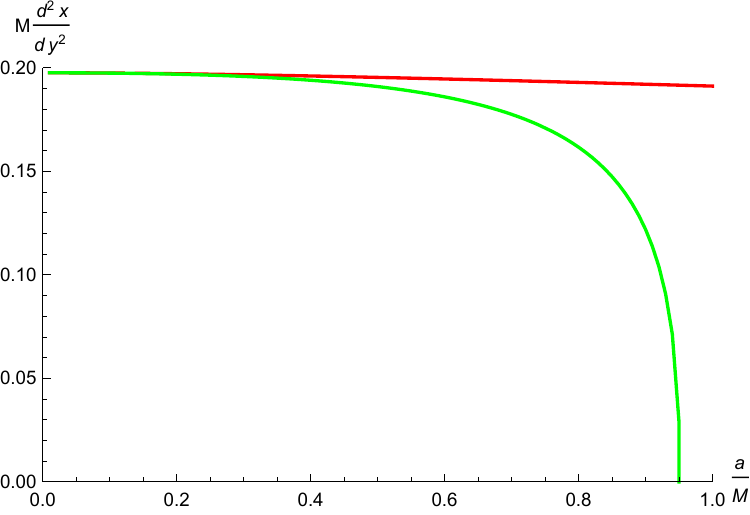}
	\qquad
	\includegraphics[width=7.5cm]{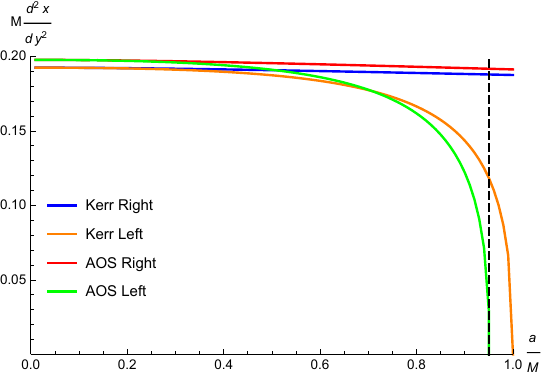}
	\caption{Comparing the curvature for the Kerr and AOS case at the right and left points of the shadow contour. The region to the left of the dashed black line indicates the allowed region for the horizon in the AOS case (see discussion in Sec IV.B.1.).}
		\label{curvfig}
	\end{figure*}	


\section{Super-radiance phenomena}\label{Super}
\subsection{Condition for super-radiance}

For super-radiance one needs to concentrate on the near horizon field modes which are radially in going. In the semi-classical level the modes can be taken as the following form (here we consider massless scalar modes)
\begin{equation}
\Psi \sim e^{iS(t,r,\theta,\phi)}~.
\label{BRM3}
\end{equation} 
In the above $S(t,r,\theta,\phi)$, for our rotating metric (\ref{BLC}) can be considered of the form (\ref{act}) with $\mu=0$.
The radial part can be determined from the solution of (\ref{Q1}). Since we are interested in near horizon regime, this equation will be solved after reducing it in the near horizon form. Note that the horizon is determined by the vanishing of $\Delta(r)$, given in (\ref{BB1}). Using the tortoise coordinate $r^*$, defined as $dr^*/dr = (k(r)+a^2)/\Delta(r)$, Eq. (\ref{Q1}) can be expressed as
\begin{eqnarray}
&&-\Big(\frac{dS_r}{dr^*}\Big)^2 + \Bigg[\frac{\Big(\sqrt{\frac{G}{F}}H + a^2\sin^2\theta\Big)E - aL}{k(r)+a^2}\Bigg]^2
\nonumber
\\
&& - \frac{(L-aE)^2+\mathcal{Q}}{(k(r)+a^2)^2}\Delta=0~.
\label{BRM4}
\end{eqnarray}
In the near horizon limit as $\Delta\to 0$, the above can be written approximately as
\begin{equation}
\Big(\frac{dS_r}{dr^*}\Big) \simeq \pm \Big[E - \frac{aL}{k(r_H)+a^2}\Big]~,
\label{BRM5}
\end{equation}
where $k(r)$ is given in (\ref{BRM1}). The solution is found out to be as
\begin{equation}
S_r \sim  \pm \Big[E - \frac{aL}{k(r_H)+a^2}\Big]r^*~.
\label{BRM6}
\end{equation}
Therefore, by (\ref{BRM3}) the near horizon mode solution comes out to be
\begin{equation}
\Psi\sim \mathcal{S}_\theta e^{-iEt}e^{iL\phi}e^{\pm i\Big[E - \frac{aL}{k(r_H)+a^2}\Big]r^*}~,
\label{BRM7}
\end{equation}
where we denote $e^{iS_\theta}$ by $\mathcal{S}_\theta$, as explicit expression of $S_{\theta}$ is unimportant for the present purpose. The value of $k(r_H)$ reduces to a very simple form:
\begin{equation}
k(r_H) = 4r_H^2\Big(1+\frac{\gamma^2L_0^2\delta_c^2r_S^2}{16r_H^4}\Big) \frac{(1+\epsilon)^{2}\big(\frac{r_H}{r_S}\big)^{2+\epsilon}}{\Big(\epsilon+(2+\epsilon)(\frac{r_H}{r_S})^{1+\epsilon}\Big)^2}~.
\label{BRM8}
\end{equation}
In (\ref{BRM7}), negative (positive) sign corresponds to the ingoing (outgoing) mode. Since we are interested for the super-radiant modes, we will concentrate on the ingoing mode solution.

In order to find the condition for ingoing mode to be super-radiant one can follow the usual procedure (e.g. see the analysis in Section $8.6$ of \cite{PaddyBook}). Since rest of the analysis is identical to the usual one, without going into the details we just give the final condition for finding super-radiant mode. Following \cite{PaddyBook} one finds the expression for energy flux through the horizon for massless scalar field as
\begin{equation}
\frac{d\mathcal{E}}{dt} = C_1 E \Big(E-\frac{aL}{k(r_H)+a^2}\Big)~,
\label{BRM9}
\end{equation}
where $C_1$ is a positive constant, whose value is not important here.
Then the condition for super-radiance in the present situation turns out to be $\Big(E-\frac{aL}{k(r_H)+a^2}\Big)<0$; i.e. 
\begin{equation}
0<E<\frac{aL}{k(r_H)+a^2}~,
\label{BRM10}
\end{equation}
where $k(r_H)$ is given by (\ref{BRM8}).

Let us now concentrate to find the angular velocity of the rotating AOS BH. This can be easily found out by considering $\theta=\pi/2$ in metric (\ref{BLC}). Under this circumstances the metric reduces to
\begin{equation}
\Big[H+a^2\Big(2\sqrt{\frac{F}{G}}-F\Big)\Big]\Big(\frac{d\phi}{dt}\Big)^2 - 2a\Big(\sqrt{\frac{F}{G}}-F\Big)\Big(\frac{d\phi}{dt}\Big)-F=0~.
\label{BRM11}
\end{equation}
Remember that in the above all the function are defined at $\theta=\pi/2$.
\begin{widetext}
The solutions are
\begin{equation}
\Big(\frac{d\phi}{dt}\Big)_{\pm}= \frac{a\Big(\sqrt{\frac{F}{G}}-F\Big) \pm\sqrt{a^2\Big(\sqrt{\frac{F}{G}}-F\Big)^2 + F\Big[H+a^2\Big(2\sqrt{\frac{F}{G}}-F\Big)\Big] }}{H+a^2\Big(2\sqrt{\frac{F}{G}}-F\Big)}~.
\label{BRM12}
\end{equation}
\end{widetext}

\begin{figure}[h]
	\centering
	\includegraphics[width=8cm]{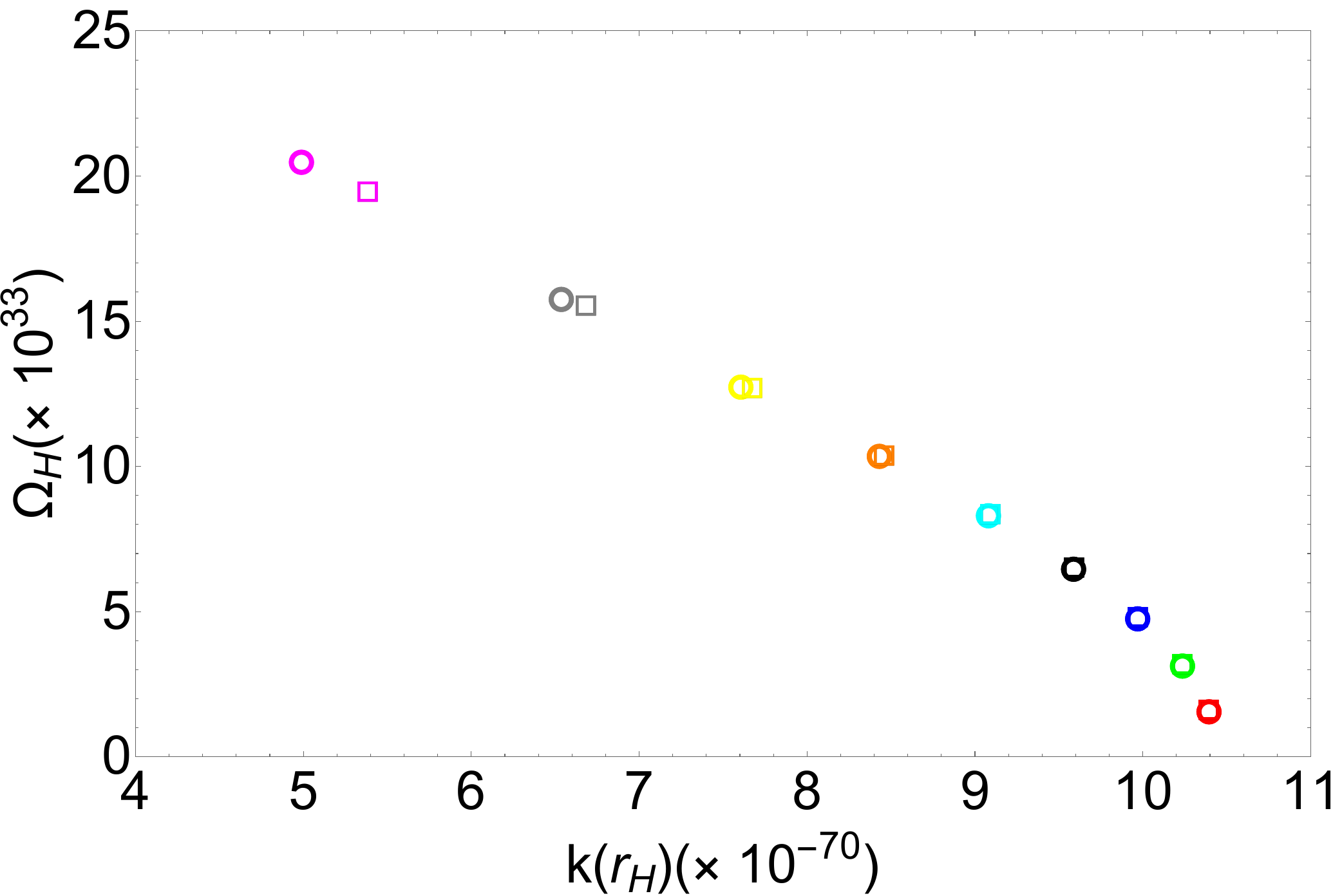}
	\caption{Behaviour of angular velocity ($\Omega_H$) of the rotating AOS and the Kerr BH with respect to $k(r_{H})$ with mass $m=1\ell_{pl}$ and for different values of the spin parameter $a$. The open circles are for the AOS BH and the open squares are for the Kerr case. Different colours of the legends imply different spin parameters for the BH, viz., Red: $a=0.1m$, Green: $a=0.2m$, Blue: $a=0.3m$, Black: $a=0.4m$, Cyan: $a=0.5m$, Orange: $a=0.6m$, Yellow: $a=0.7m$, Grey: $a=0.8m$ and Magenta: $a=0.9m$ respectively. For smaller values of $a$, there is an overlap between the rotating AOS and Kerr results.}
	\label{fig:avel}
\end{figure}

The horizon angular velocity is given by the minimum value of $\Big(\frac{d\phi}{dt}\Big)$, calculated at the horizon. For $\theta = \pi/2$, from (\ref{F}) and (\ref{G}) we have $F = (gh/k^2)H$ and $G = gh/H$. Substituting these in the negative sign solution in (\ref{BRM12}) one finds
\begin{equation}
\Big(\frac{d\phi}{dt}\Big)_- = \frac{a\Big(\frac{1}{k} - \frac{gh}{k^2}\Big) - \frac{1}{k}\sqrt{a^2+gh}}{1+ a^2\Big(\frac{2}{k} - \frac{gh}{k^2}\Big)}~.
\label{BRM13}
\end{equation} 
Now at the horizon we have $\Delta(r_H) = 0$ which, from (\ref{BB1}), yields $g(r_H)h(r_H) + a^2=0$. Using this in the above we find the angular velocity of the BH as
\begin{equation}
\Omega_H = \frac{a}{k(r_H) + a^2}~.
\label{BRM14}
\end{equation}
Then in terms of the angular velocity the condition for super-radiance take the following form:
\begin{equation}
0<E<L\Omega_H~,
\label{BRM15}
\end{equation}
which is similar in form for the usual Kerr BH. In this case only $\Omega_H$ has been modified which incorporates all the quantum effects. A behaviour of the angular velocity of the horizon with respect to $k(r_{H})$ for the AOS BH with mass $1\ell_{pl}$ and for different values of the spin parameters (starting from $a=0.1 m$ to $a=0.9m$ is plotted in Fig. (\ref{fig:avel}). We have also plotted the respective Kerr angular velocity in the same plot for the same value of mass. It turns out that the behaviour of the horizon angular velocity of the Kerr BHs and the rotating AOS BHs are very similar and almost exactly matches for small values of the spin parameter $a$, while, as the spin parameter increases, the value of the angular velocity starts to change significantly. 

\vspace{100 px}

\subsection{Amplification factor: scalar field scattering and the Teukolsky formalism}

The calculation of superradiance amplification factors can be done using the Teukolsky formalism. The approach is discussed in detail in the recent review \cite{brito2020superradiance}. In the Kerr geometry, the perturbations can be written down in the form of the Teukolsky master equation \cite{teukolsky1972rotating,teukolsky1973perturbations}. 
For scalar perturbation, the dynamical equation is given by the Klein Gordon equation in this curved spacetime. Here we consider massless scalar fields $\psi$ propagating in the rotating AOS background (\ref{BLC}). Thus, the Klein Gordon equation is given by $\Box \psi = 0$. Using the symmetry in $(t,\phi)$, we consider the ansatz
\begin{equation}\label{ansatz}
\psi(t,r,\theta,\phi) = e^{-i\left(\omega t + q \phi \right)}S(\theta) J(r)~,
\end{equation}
where $q$ is the azimuthal number. Under that above form of solution, in tortoise coordinate we can then write the radial part of the Klein Gordon equation in the following Schrödinger like form (see Appendix \ref{RadialEqnSol}): 
\begin{equation}\label{1Abhinove}
\dv[2]{\Phi}{r_*} + V_{\text{eff }} \Phi = 0~,
\end{equation}
where function $\Phi$ is related to the radial function $J(r)$ as $\Phi = \sqrt{k(r) + a^2}J$. The form of $V_{\textrm{eff}}$ is given in Appendix \ref{RadialEqnSol}. The solutions of the above equation is obtained under a boundary condition such that their asymptotic form must satisfy
\begin{equation}\label{2Abhinove}
\Phi = \begin{cases}
	T e^{-i k_H r_*} & \text{for } r \to r_H \\
	Ie^{-ik_{\infty} r_*} + Re^{i k_{\infty} r_*} & \text{for } r \to \infty~,
	\end{cases}
\end{equation}
the wave numbers $k_H$ and $k_{\infty}$ are given by $k_H^2 = V_{\text{eff}}(r \rightarrow r_H)$ and $k_{\infty}^2 = V_{\text{eff}}(r \rightarrow \infty)$, respectively. 
Moreover the relation between different coefficients is given by \cite{brito2020superradiance}
\begin{equation}\label{3Abhinove}
|R|^2 = |I|^2 - \frac{k_H}{k_{\infty}}|T|^2~.
\end{equation}
For superradiant amplification, we must have $\abs{R}^2 > \abs{I}^2$. It will be satisfied if one has $\frac{k_H}{k_{\infty}} < 0$.
For scalar fields, the amplification factor is defined as \cite{brito2020superradiance}
\begin{equation}\label{ampfact}
    Z_{slq} = Z_{0lq} = \abs{\frac{R}{I}}^2 - 1~.
\end{equation}
Using the above one we will find the amplification factor for AOS black hole.


To obtain the coefficients $R,I$, we need to solve the radial part (\ref{1Abhinove}). 
In terms of $J$ variable and $r$ coordinate this takes the form
\begin{equation}\label{5Abhinove}
    \dv{}{r}\left({\Delta \dv{J}{r}}\right) + \left(\frac{K^2}{\Delta} - \lambda \right)J = 0~, 
\end{equation}
where $K = (k(r)+a^2)\omega - qa$ and $\lambda = A_{0lq} + a^2\omega^2 - 2 a q \omega$ is the constant of separation. $A_{0lq}$ are the angular eigenvalues (see Appendix \ref{RadialEqnSol}). To solve it we consider the small rotation approximation, as adopted in  \cite{brito2020superradiance}. For sufficiently small rotation i.e $a\omega << 1$, the angular eigenvalues are given by $A_{0lq} = l(l+1) + \mathcal{O}(a^2\o^2)$. In what follows, we set $\hbar = 1$ (in addition to $G = c = 1$) and take $E = \omega, L = q$ and take $m=GM/c^2 \equiv M$ as the black hole mass. Note that solving (\ref{5Abhinove}) is equivalent to solving (\ref{1Abhinove}) since one can derive (\ref{1Abhinove}) from (\ref{5Abhinove}) as shown in Appendix \ref{RadialEqnSol}. 

 We solve this using matching asymptotic techniques, which is discussed in detail in Appendix (\ref{RadialEqnSol}). It must be noted to obtain the solution one should have the knowledge of the outer horizon. 
 In the case of AOS, we have not yet considered the structure of the horizon. But in the regime where quantum corrections are small, we demand that the outer horizon of AOS be obtained by finding the quantum corrections to the outer horizon of Kerr. In the following section, we discuss in detail how to obtain these roots, upto first order in $\epsilon$.

\subsubsection{Horizon Structure}

For the rotating AOS black hole, we do not yet have an analytical expression for the horizon. This is difficult since the roots of $\Delta =0$ generally give the horizon, but here, the quantum parameters are found in the exponent of the variable $r$ in the function, if one explicitly writes it down. In this context, we consider ``small'' quantum corrections such that one can expand $\Delta$ about the Kerr outer horizon. That is to say, we will consider a Taylor's expansion of $\Delta$ upto 2nd order about $r = r_+^{\text{Kerr}}$ and up to first order about $\epsilon = 0$, which gives us two roots for $\Delta$. On considering the classical limit, we find that one of the roots reduces to $r_+^{\text{Kerr}}$. We denote this root as $r_+$ in what follows, and is (up to first order in $\epsilon$) the outer horizon of the rotating AOS metric. The other root represents some other surface which does not have the interpretation of the horizon, and we shall denote this by $r^{\prime{}}$. Thus, around $r = r_+^{\text{Kerr}}$, we can approximately write $\Delta$ in the form $\Delta = \left(r-r_+\right)\left(r-r^{\prime{}}\right)$, upto first order in $\epsilon$. The actual expressions are included in a Mathematica notebook \cite{Mathematica}. 

\begin{figure}[!h]
    \centering
    \begin{minipage}{0.5\textwidth}
        \centering
        \includegraphics[width=0.92\linewidth]{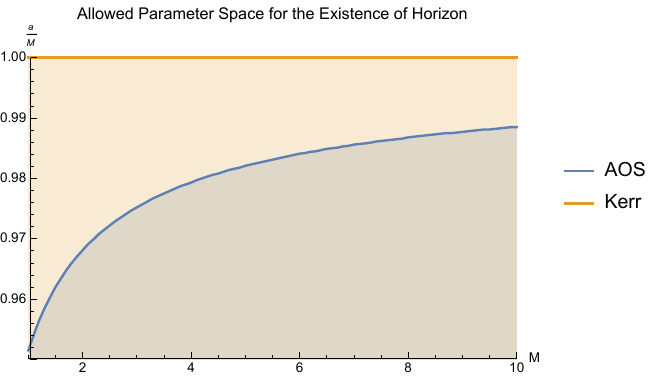}
    \end{minipage}
    \begin{minipage}{0.5\textwidth}
        \centering
        \includegraphics[width=0.92\linewidth]{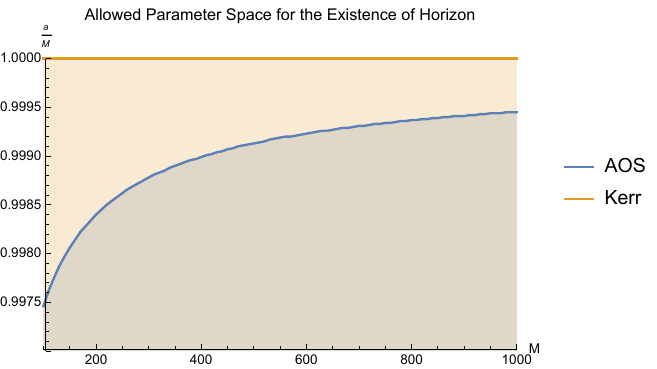}
        \caption{Comparing allowed parameter space in $\f{a}{M}$ vs $M$ plane \\ for Kerr and AOS.}\label{newfig2}
    \end{minipage}
\end{figure}



Usually, demanding that $\Delta = 0$ has real roots is used to find the location of the horizon and to avoid a naked singularity. For the Kerr black hole, this is achieved by the condition $M>a$. In the AOS case, this is modified. Given the form of $\Delta$, one can check whether roots to this equation exist, in the region where the classical condition holds. That is, we want to check the horizon structure of the rotating AOS black hole, inside the classically allowed region. In doing so, we find the following parameter space. We see that even with maintaining $M>a$, certain values of $a$ are only available to black holes with higher masses since $\Delta$ becomes completely positive and hence has no roots, when $\f{a}{m}$ is high. The precise threshold of $\f{a}{m}$ past which $\Delta$ fails to have roots depends on the value of $M$ and is seen to increase, for increasing $M$. This can be observed from the parameter space plot, shown in Fig (\ref{newfig2}). The complete region represents the classically allowed region of parameters. The blue curve and the region under the curve represents the threshold at which the rotating AOS black hole horizon ceases to exist since $\Delta$ ceases to have roots beyond the threshold value of $\f{a}{M}$. Therefore, even in the classically allowed region, the quantum corrections play a role so as to further restrict the parameter space under which the rotating AOS black hole horizon is well defined. Note that this does not require the first order expansion in $\epsilon , \delta$ since this is obtained directly from the function $\Delta$. A simulation showing the change in the function $\Delta$ is included in the Mathematica notebook attached \cite{Mathematica}.

\subsubsection{Estimation of amplification factor}
\begin{figure}[h!]
    \centering
    \begin{subfigure}{}
        \centering
        \includegraphics[width=0.5\textwidth]{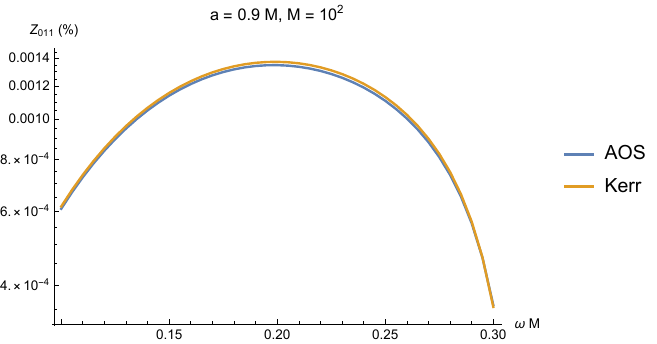}
    \end{subfigure}%
    ~ 
    \begin{subfigure}{}
        \centering
        \includegraphics[width=0.5\textwidth]{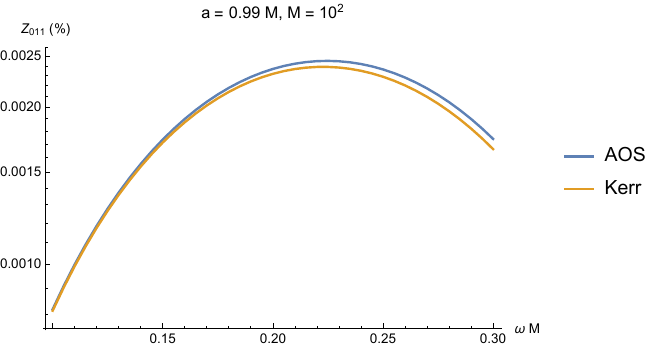}
    \end{subfigure}
    ~
    \begin{subfigure}{}
        \centering
        \includegraphics[width=0.5\textwidth]{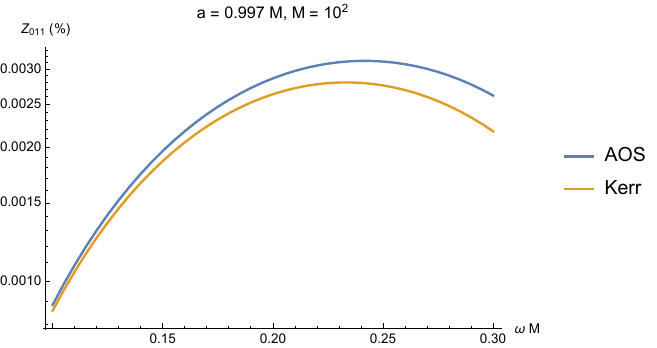}
    \end{subfigure}
    \caption{Comparing the effect of the rotation parameter $a$ for $Z_{011}$ with $M  = 10^2$ (in units of $l_{Pl}$), We see that for black holes of a constant mass, an increase in the rotation parameter increases the separation between the AOS and Kerr superradiance. In this mass regime, superradiant amplification in AOS starts out lower than that of Kerr, but increases and exceeds Kerr with an increase in $a$}\label{Fig3Abhinove}
\end{figure}

\begin{figure}[h!]
    \centering
    \begin{subfigure}{}
        \centering
        \includegraphics[width=0.5\textwidth]{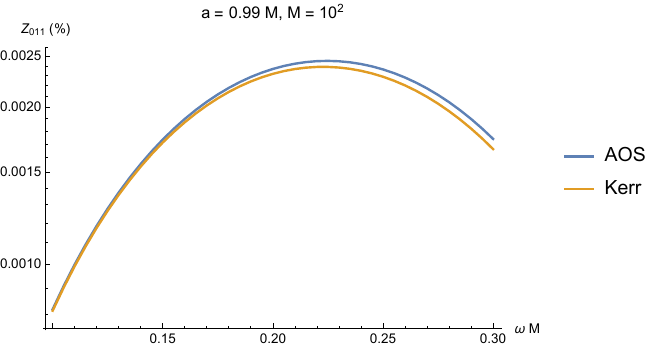}
    \end{subfigure}%
    ~ 
    \begin{subfigure}{}
        \centering
        \includegraphics[width=0.5\textwidth]{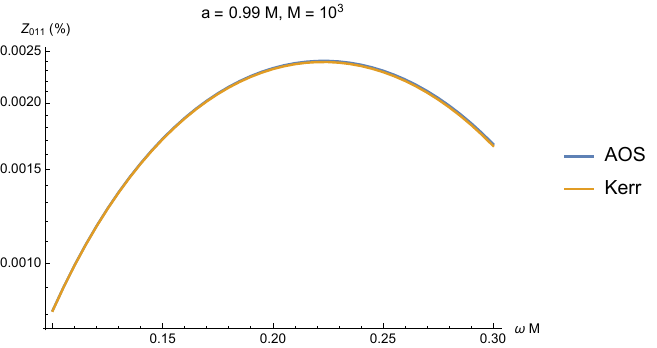}
    \end{subfigure}
    ~
    \begin{subfigure}{}
        \centering
        \includegraphics[width=0.5\textwidth]{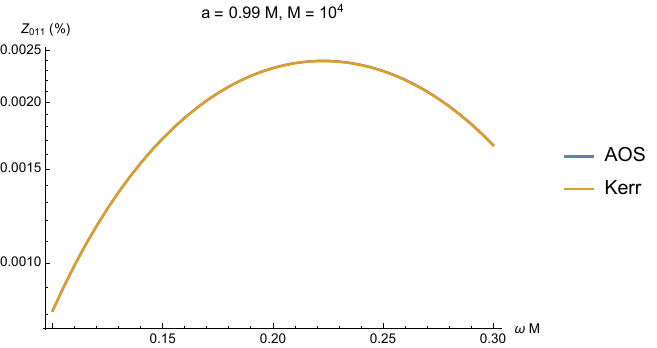}
    \end{subfigure}
    \caption{Comparing the effect of increasing mass $M$ on $Z_{011}$ with $a  = 0.99 M$ (with $M$ in units of $l_{Pl}$). By varying the mass with a constant rotation parameter, we see the opposite effect.  The amplification is much more closer between Kerr and AOS. This can be understood by noting that an increase in mass corresponds to a decrease in quantum correction effects. Thus, the large mass limit and the classical limit, are identical.}\label{Fig4Abhinove}
\end{figure}

We can now solve the radial equation, (see Appendix \ref{RadialEqnSol}) and obtain the amplification factors. The amplification factors $Z_{011}$ and $Z_{033}$ are plotted as a function of $\omega M$ in Fig \ref{Fig3Abhinove} and Fig \ref{Fig4Abhinove}. We look at the effect of the rotation parameter $a$, effect of the mass of the black hole 
in Fig \ref{Fig3Abhinove} and\ref{Fig4Abhinove} respectively. 




\section{Conclusion}\label{Con}

We investigated the properties of the shadow of a recently obtained regular BH in \cite{Akiyama:2019cqa,Akiyama:2019fyp,Akiyama:2019eap}. The spacetime is being modified from the LQG inspired theory. It is found that the singularity $r=0$ is now hidden by a lowest possible value of the area element. We studied shadows for both non-rotating and as well rotating cases. It is being found that the quantum corrections can put significant signature in shadow when the BH size is of the Planck order. We noticed that although the shape of the shadows, comparing with those for Schwarzschild BH, do not change in non-rotating case in presence of the quantum corrections, but the area of the shadow decreases. The shadow radius increases as we increase mass of the BH and after certain value of mass both are indistinguishable. This shows that the quantum correction makes prominent signature on shadow for a very small Planck mass BH.

The rotating counterpart also behaves similarly. Here also the area of the shadow decreases due to quantum effect. Moreover, the contraction is less on the left hand side compared to the other side in rotating AOS BH shadow when compared with its classical counterpart, i.e. the Kerr BH. As we increase the mass of the BH, the distortion from Kerr starts to reduce. Increase of the rotation also imparts more distortion on the right side while the same decreases on the left part.  

We also studied the super-radiance phenomenon for the rotating AOS BH. We observed that the condition for massless scalar field super-radiance is identical to that of the Kerr case. But here, the rotation of the BH is more than that compared to Kerr in the low mass regime. As we increase mass of BH the difference from Kerr starts to become insignificant. This implies that for low mass AOS BH the window for energy of scalar field to perform super-radiance is larger than higher mass system. As we increase mass, this window will become narrower and ultimately for large mass the window will coincide with Kerr value. The amplification factors themselves also exhibit this trend, as seen in Fig \ref{Fig3Abhinove} and Fig \ref{Fig4Abhinove} where for larger black holes, the scalar field amplification coincides for AOS and Kerr. A natural next step would be to understand superradiant scattering of other spin fields (in this work we only consider $s = 0$) in this background. This can be done by looking at the perturbations using the Newman Penrose formalism \cite{newman1962approach}, as done by Teukolsky, Press and others \cite{teukolsky1972rotating},\cite{teukolsky1973perturbations},\cite{press1973perturbations} for the Kerr geometry. One can make a straightforward generalisation of the Kinnersely tetrad in the Kerr case, to look at other spin fields in AOS. 

Finally, it may be worth mentioning that the primordial black holes (PBH) are microscopic in size and therefore the quantum effects at this scale are not going to be negligible. Since AOS BH shows significant quantum effects at the Planck scale it may model these PBHs. Therefore investigation of these quantum induced spacetime can be useful to understand the nature of PBH. It is our belief that the present study will illuminate the properties of Planck scale physics. 


\vskip 3mm
\noindent
{\bf Acknowledgement:} The research of BRM is partially supported by a STARTUP RESEARCH GRANT (No. SG/PHY/P/BRM/01)
from the Indian Institute of Technology Guwahati, India.

\begin{widetext}
\begin{appendix}
\section{Derivation of metric for rotating BH through Newman-Janis algorithm (NJA)}\label{Appendix A}
\subsection{Original NJA}\label{AppA1}

The NJA, originally proposed in  \cite{Newman:1965tw,Newman:1965my}, helps to  construct  a stationary and axisymmetric metric from a static and spherically symmetric metric of the form
\begin{equation}
ds^2=-f(r) dt^2+\frac{dr^2}{g(r)}+h(r)\left(d\theta^2+\sin^2\theta d\phi^2\right)~.
\label{sss}
\end{equation}
To find the required result there are certain prescribed steps to be followed. Without going into the details, let us here just follow the prescribed steps.

The {\it first step} of the algorithm is to write down the metric (\ref{sss}) in the advance null
(Eddington-Finkelstein) coordinates $(u,r,\theta,\phi)$. This is achieved by using the following transformation:
\begin{equation}
du=dt-\frac{dr}{\sqrt{fg}}~.
\end{equation}
The non rotating metric in the advance null coordinates then becomes
\begin{equation}
ds^2=-f(r) du^2-2\sqrt{\frac{f}{g}}dudr+h(r)\left(d\theta^2+\sin^2\theta d\phi^2\right)~.
\end{equation}
The {\it second step} is to use a null tetrad $Z_\alpha^\mu=(l^\mu,n^\mu,m^\mu,\bar{m}^\mu)$  to express the inverse metric $g^{\mu\nu}$ in the form
\begin{equation}
g^{\mu\nu}=-l^\mu n^\nu -l^\nu n^\mu +m^\mu \bar{m}^\nu +m^\nu \bar{m}^\mu~,
\end{equation}
Here, the tetrad vectors satisfy the relations
\begin{eqnarray}
&&l_\mu l^\mu = n_\mu n^\mu = m_\mu m^\mu = l_\mu m^\mu = n_\mu m^\mu =0~;
\nonumber
\\
&&l_\mu n^\mu = - m_\mu \bar{m}^\mu =-1~.
\end{eqnarray}
and $\bar{m}^\mu$ is the complex conjugate of $m^\mu$.

The tetrad vectors satisfying the above relations are found to be
\begin{equation}
l^\mu=\delta^\mu_r, \hspace{0.2cm} n^\mu=\sqrt{\frac{g}{f}}\delta^\mu_u-\frac{g}{2}\delta^\mu_r, \quad m^\mu=\frac{1}{\sqrt{2h}}\left(\delta^\mu_\theta+\frac{i}{\sin\theta}\delta^\mu_\phi\right)~.
\label{TV}
\end{equation}
The {\it third step} is to perform  the complex coordinate transformations in the $(u, r)$ plane,
\begin{equation}
u'= u-i a \cos\theta; \,\,\,\ r'= r+i a \cos\theta~,
\label{com}
\end{equation}
where $a=\f{J}{m}$ will be identified as the specific angular momentum or spin parameter of the BH, and $m$, $J$ are the mass and angular
momentum of the black hole, respectively.
Under these transformations the new null tetrads are
\begin{equation}
l'^\mu=\delta^\mu_r, \quad n'^\mu=\sqrt{\frac{G(r,\theta)}{F(r,\theta)}}\delta^\mu_u-\frac{G(r,\theta)}{2}\delta^\mu_r; \,\,\,\
m'^\mu=\frac{1}{\sqrt{2H(r,\theta)}}\left(ia\sin\theta(\delta^\mu_u-\delta^\mu_r)+\delta^\mu_\theta+\frac{i}{\sin\theta}\delta^\mu_\phi\right)~,
\label{BRM2}
\end{equation}
where $F(r,\theta)=f(r')$, $G(r,\theta) = g(r')$ and $H(r,\theta) = h(r')$ are, respectively, the complexified form of $f(r)$, $g(r)$ and $h(r)$. 
Using the new tetrad, the new inverse metric is found to be
\begin{equation}
g'^{\mu\nu}=-l'^\mu n'^\nu -l'^\nu n'^\mu +m'^\mu \bar{m}'^\nu +m'^\nu \bar{m}'^\mu~.
\end{equation}
Then the new metric in the advance null coordinates becomes
\begin{eqnarray}
ds^2&=&-Fdu^2-2\sqrt{\frac{F}{G}}dudr+2a\sin^2\theta\left(F-\sqrt{\frac{F}{G}}\right)du d\phi+2a\sqrt{\frac{F}{G}}\sin^2\theta drd\phi +H d\theta^2
\nonumber
\\
&+&\sin^2\theta\left[H+a^2\sin^2\theta\left(2\sqrt{\frac{F}{G}}-F\right)\right]d\phi^2~.
\label{efc}
\end{eqnarray}

The {\it final step} of the algorithm is to rewrite the above metric in Boyer-Lindquist form (where the only nonzero off diagonal term is $g_{t'\phi'}$) using the global coordinate transformations of the form
\begin{equation}
du=dt{'}+\chi_{1}(r)dr,\quad \quad d\phi=d\phi^{'}+\chi_{2}(r)dr~. \label{GT}
\end{equation}
In the above $\chi_1(r)$ and $\chi_2(r)$ are chosen in such a way that the metric will have only $g_{t'\phi'}$ non-vanishing off-diagonal term in $(t',r,\theta,\phi')$ coordinates. Substitution of (\ref{GT}) in (\ref{efc}) and then demanding the above criterion one finds the metric of the form (\ref{BLC}) 
with the following choices of $\chi_1(r)$ and $\chi_2(r)$:
\begin{eqnarray}
	\chi_{1}(r)&&=-\frac{\sqrt{\frac{G(r,\theta)}{F(r,\theta)}}H(r,\theta)+a^{2}\sin^{2}\theta}{G(r,\theta)H(r,\theta)+a^{2}\sin^{2}\theta}\equiv -\frac{A(r)}{B(r)}~, 
	\label{chi1}
	\\
	\chi_{2}(r)&&=-\f{a}{G(r,\theta)H(r,\theta)+a^{2}\sin^{2}\theta} \equiv -\frac{a}{B(r)}~.
	\label{chi2}
\end{eqnarray}
In (\ref{BLC}) we dropped the prime in time and azimuthal coordinates. This line element represents the desired metric for stationary, axisymmetric spacetime.

It must be mentioned that the transformations in (\ref{GT}) are possible (i.e. these have to be integrable) only when $\chi_{1}(r)$ and $\chi_{2}(r)$ are functions of only r and not $\theta$. This implies that the denominator in (\ref{chi2}) must be function of $r$ only, which we call as $B(r)$. Consequently, the numerator in (\ref{chi1}) must be again function of $r$ only (we call this as $A(r)$). This is a very non-trivial restriction and may not be always satisfied for any value of $f(r), g(r)$ and $h(r)$ under the complexification (\ref{com}). Incidentally for Schwarzschild black hole this is satisfied and one obtains the Kerr metric from (\ref{BLC}). Whereas our present static, spherically symmetric metric coefficients, given in (\ref{fr}), (\ref{gr}) and (\ref{hr}), do not satisfy these conditions. Therefore the above ditto procedure fails to provide the rotating solution for AOS black hole. 

\subsection{Modified NJA}\label{AppA2}
We see that the original NJA may fail for some spherically symmetric static metric, like the present one. A little modification in the approach, as shown in \cite{PhysRevD.90.064041}, successfully overcomes this difficulty. For that it is assumed that we have somehow obtained a metric of the form (\ref{efc}). In this case we do not know the exact form of $G, F$ and $H$; i.e. we are not using the specific complexification, given in (\ref{com}). But the null tetrads are of the form (\ref{BRM2}) so that we have a rotating metric, given by (\ref{efc}). Then in the {\it final step} of the earlier subsection, take again the transformations of $u$ and $\phi$ coordinates similar in form as (\ref{GT}): 
\begin{equation}
du=dt{'}+\lambda(r)dr,\quad \quad d\phi=d\phi^{'}+\chi(r)dr~. 
\label{GT1}
\end{equation}
But here choose the unknown functions as
\begin{equation}
\lambda(r)=-\f{k(r)+a^{2}}{g(r)h(r)+a^{2}}~;\,\,\,\
\chi(r)=-\f{a}{g(r)h(r)+a^{2}}~;\,\,\
k(r)=\sqrt{\f{g(r)}{f(r)}}h(r)~.
\label{BRM1}
\end{equation}
Note that the above choice is inspired by the forms (\ref{chi1}) and (\ref{chi2}). Since $\lambda$ and $\chi$ have to be function of $r$ only, the original metric coefficients of (\ref{sss}) are intentionally positioned at the places where, in (\ref{chi1}) and (\ref{chi2}), we had the complexified versions of them. Also $\sin^2\theta$ has been removed. This guarantees only radial dependence of our unknown functions for any given static, spherically symmetric metric. 

In order to find $F, G$ and $H$ we will use the earlier trick. Since we want metric to be in Boyer-Lindquist form, 
after inserting (\ref{GT1}) with (\ref{BRM1}) in (\ref{efc}), set $g_{t'r} $ and $ g_{r\phi^{'}}$ to be zero. This will give us the relations among $F, G, H$ and known functions $f(r), g(r), h(r)$. But since we have only two equations corresponding to vanishing of two off-diagonal metric coefficients, one function among $F, G, H$ will remain undetermined. This yields
\begin{eqnarray}
&&F=\f{g(r)h(r)+a^{2}\cos^{2}\theta}{(k(r)+a^{2}\cos^{2}\theta)^{2}}H~;
\label{F}
\\
&&G=\f{g(r)h(r)+a^{2}\cos^{2}\theta}{H}~,
\label{G}
\end{eqnarray}
where we choose $H$ to be undetermined.
In this case the metric (\ref{efc}) takes the form
\begin{equation}
ds^{2}=-Fdt^{2}-2a\sin^{2}\theta\bigg(\sqrt{\f{F}{G}}-F\bigg)dt d\phi+\f{H}{g(r)h(r)+a^{2} }dr^{2}
+H d\theta^{2}+\sin^{2}\theta \bigg[H+a^{2}\sin^{2}\theta\bigg(2\sqrt{\f{F}{G}}-F\bigg)\bigg]d\phi^{2}~,
\label{BLC3}
\end{equation}
where we have dropped the primes in t and $\phi$.
Now as from (\ref{G}) we have
\begin{equation}
GH+a^{2}\sin^{2}\theta=g(r)h(r)+a^{2}~,
\end{equation}
the metric (\ref{BLC3}) reduces to our desired form (\ref{BLC}). Remember that in this case $H$ remains undetermined, whereas $F$ and $G$ are given by (\ref{F}) and (\ref{G}), respectively.

\section{Photon trajectories}\label{Appendix B}
The Hamilton-Jacobi (H-J) equation is given by
\begin{equation}
\f{\p S}{\p \l}+\f{1}{2}g^{\mu\nu}p_{\mu}p_{\nu}=0~,
\label{HJ1}
\end{equation}
where $S$ is the Jacobi action of the photon, $\l$ is the affine parameter of the null geodesic and $p_{\mu}$ is the momentum given by
\begin{equation}
p_{\mu}=\f{\p S}{\p x^{\mu}}=g_{\mu\nu}\f{dx^{\nu}}{d\l}~.
\label{mom}
\end{equation}
Now from (\ref{BLC}) one can see that the metric $g_{\mu\nu}$ is independent of $t$ and $\phi$. Therefore we have two constants of motion --
the conserved energy ($E$) and the angular momentum of the photon in the direction of the rotation axis ($L$).
In that case the ansatz for $S$ is taken as
\begin{equation}
S=\f{1}{2}\mu^{2}\l -Et+L\phi +S_{r}(r)+S_{\theta}(\theta)~, 
\label{act}
\end{equation}
where $\mu$ is the rest mass of the particle  moving in the black hole spacetime. For photons we have $\mu=0$. In the above choice the radial and $\theta$ dependence are taken to be separated as we have another conserved quantity, known as Carter constant, for our metric. 
Putting Eq. (\ref{act}) in the Hamilton-Jacobi equation (\ref{HJ1}), we obtain after some simplifications
\begin{eqnarray}
-\left(GH+a^2\sin^2\theta\right)& &\left(\frac{dS_r}{dr}\right)^2+\frac{\left[\left(\sqrt{\frac{G}{F}}H+a^2\sin^2\theta\right)E-aL \right]^2}{\left(GH+a^2\sin^2\theta\right)}-(L-aE)^2 \no\\
&&=\left(\frac{dS_\theta}{d\theta}\right)^2+L^2\cot^2\theta-a^2E^2\cos^2\theta~. 
\label{HJ3}
\label{rtheta}
\end{eqnarray}
Now, since the quantities $\left(GH+a^2\sin^2\theta\right)=g(r)h(r)+a^{2}=\Delta(r)$ and
$\left(\sqrt{\frac{G}{F}}H+a^2\sin^2\theta\right)=k(r)+a^{2}=\sqrt{\f{g(r)}{f(r)}}h(r)+a^{2}=\Sigma(r)$ are functions of $r$ only, the left- and right-hand side of
Eq. (\ref{HJ3}) are only functions of $r$ and $\theta$, respectively. Therefore, each side of this equation must be equal to some separation constant. Thus introducing the Carter constant $\mathcal{Q}$ as the separation constant we have after separation
\begin{equation}
-\left(GH+a^2\sin^2\theta\right)\left(\frac{dS_r}{dr}\right)^2+\frac{\left[\left(\sqrt{\frac{G}{F}}H+a^2\sin^2\theta\right)E-aL \right]^2}{\left(GH+a^2\sin^2\theta\right)}-(L-aE)^2= \mathcal{Q}~,
\label{Q1}
\end{equation}
and
\begin{equation}
\left(\frac{dS_\theta}{d\theta}\right)^2+L^2\cot^2\theta-a^2E^2\cos^2\theta=\mathcal{Q}~. 
\label{Q2}
\end{equation}
The geodesic equations (\ref{eq:t_eqn}), (\ref{eq:phi_eqn}), (\ref{eq:r_eqn}) and (\ref{eq:theta_eqn})  are
then obtained from (\ref{mom})  by making use of (\ref{Q1}) and (\ref{Q2}).

\section{Klein Gordon Equation in Rotating AOS Spacetime}\label{RadialEqnSol}

\subsection{Deriving the Schrödinger like form}
We want to solve $\Box \psi = 0$.
From the line element (\ref{BLC}), the inverse metric whose elements are given by: 
\begin{align}
    g^{tt} &= -\frac{\rho^4 + 2a^2 \sin^2{\theta} \rho^2 + a^4 \sin{\theta}^4 - a^2 \sin^2{\theta}}{\Delta H} ~;\\
    g^{t\phi} &= \frac{a\Delta - a^3 \sin^2{\theta} - a\rho^2}{\Delta H}~; \\
    g^{rr} &= \frac{\Delta}{H}~; \\
    g^{\theta \theta} &= \frac{1}{H}~; \\
    g^{\phi \phi} &= \frac{\Delta - a^2 \sin^2{\theta}}{\Delta H \sin^2{\theta}}~.
\end{align}

Thus, expanding we have the equation: 
\begin{multline}\label{AppAbhinove2}
    \left(\frac{(k+a^2)^2 - \Delta a^2 \sin^2{\theta}}{\Delta} \right)\pdv[2]{\psi}{t} - \frac{2a\left(gh - k\right)}{\Delta}\frac{\partial^2 \psi}{\partial t \partial \phi} - \left(\frac{\Delta - a^2\sin^2{\theta}}{\Delta \sin^2{\theta}}\right)\pdv[2]{\psi}{\phi} \\ - \frac{\rho^2}{H}\frac{\partial}{\partial r}\left(\frac{H \Delta}{\rho^2} \pdv{\psi}{r}\right) - \frac{\rho^2}{H \sin{\theta}}\frac{\partial}{\partial \theta}\left(\frac{H \sin{\theta}}{\rho^2}\pdv{\psi}{\theta}\right) = 0~.
\end{multline}
Note that the form of $H$ is unknown to us. Without this the solution can not be obtained. We proceed in the following way to encounter the situation.

Under the classical limit the functions $k,f,g,h$ have the limits given by: $h(r) \to r^2$, $\sqrt{\frac{g(r)}{f(r)}} \to 1$, $k(r) \to r^2$. 
Thus, if we take the classical limit of (\ref{AppAbhinove2}), demanding that this must be the Klein Gordon equation for a massless scalar field in Kerr, we can fix the form for the so far undetermined function $H$. This turns out to be $H = \rho^2 = k(r) + a^2\cos^2{\theta}$. Substituting this into the Klein Gordon equation, we have: 
\begin{multline}\label{AppAbhinove3}
    \left(\frac{(k+a^2)^2 - \Delta a^2 \sin^2{\theta}}{\Delta} \right)\pdv[2]{\psi}{t} - \frac{2a\left(gh - k\right)}{\Delta}\frac{\partial^2 \psi}{\partial t \partial \phi} - \left(\frac{\Delta - a^2\sin^2{\theta}}{\Delta \sin^2{\theta}}\right)\pdv[2]{\psi}{\phi} \\ - \frac{\partial}{\partial r}\left( \Delta \pdv{\psi}{r}\right) - \frac{1}{ \sin{\theta}}\frac{\partial}{\partial \theta}\left(\sin{\theta}\pdv{\psi}{\theta}\right) = 0~.
\end{multline}
Now, we consider the ansatz given in (\ref{ansatz}).
Using this in the above equation we can separate (\ref{AppAbhinove3}), and obtain the following equations:
\begin{equation}\label{AppAbhinove4}
    \frac{d}{dr}\left(\Delta \dv{J}{r}\right) +\left(\frac{\omega^2 (k+a^2)^2 + 2 q \omega a\left(gh - k\right) + q^2 a^2}{\Delta} - a^2 \omega^2 - A_{0lq} \right) J = 0~;
\end{equation}
and 
\begin{equation}\label{AppAbhinove5}
    \frac{1}{ \sin{\theta}}\frac{d}{d \theta}\left(\sin{\theta}\dv{S}{\theta}\right) + \left(\omega^2 a^2 \cos^2{\theta} - \frac{q^2}{\sin^2{\theta}} + A_{0lq} \right)S = 0~,
\end{equation}
where $\lambda = A_{0lq} + a^2\o^2 - 2 a q \o$ is the constant of separation. 
Comparing the angular equation (\ref{AppAbhinove5}) with what one would obtain in Kerr, we can see that they are identical. In fact, this is to be expected since the quantum corrections are only found in the functions of the metric that depend on $r$ and do not have any pure angular dependence. Consequently, the solution to $S(\th)$ is the spin weighted spheroidal harmonics (with spin $s=0$ for scalar fields). 



The radial equation (\ref{AppAbhinove4}) is rewritten as
\begin{equation}\label{AppAbhinove6}
    \dv{}{r}\left({\Delta \dv{J}{r}}\right) + \left(\frac{K^2}{\Delta} - \lambda \right)J = 0~, 
\end{equation}
where $K = (k(r)+a^2)\omega - ma$ and $H = \rho^2$. We solve this using matching asymptotic techniques, which is explained below. First, we write it in a Schrödinger like form in the tortoise coordinate: 
\begin{equation}\label{tortoise}
    \dv{r}{r_*} = \frac{\Delta}{k(r) + a^2} \equiv \mu(r)~.
\end{equation}
Using this, and rewriting (\ref{AppAbhinove6}), we have: 
\begin{equation}
    \dv[2]{J}{r_*} + \frac{k^{\prime} \Delta}{(k+a^2)^2}\dv{J}{r_*} + \left(\frac{K^2 - \Delta(\lambda + a^2\omega^2 - 2aq\omega)}{(k(r)+a^2)^2}\right)J = 0~. 
\end{equation}
Now, we take a transformation to define a new function: $\Phi = \sqrt{k+a^2}J$. Using this, we can write down the final Schrödinger like equation as: 
\begin{equation}\label{AppAbhinove7}
    \dv[2]{\Phi}{r_*} + V_{\text{eff }} \Phi = 0~,
\end{equation}
where the effective potential is given by: 
\begin{equation}
    V_{\text{eff} } = \left(\frac{K^2 - \Delta (\lambda + a^2\omega^2 - 2aq\omega)}{(k+a^2)^2} - u(r)^2 - \mu u^{\prime}(r) \right)~.
\end{equation}
which is implicitly a function of $r_*$ and $u(r) = \frac{\mu k^{\prime}}{2(k+a^2)^2}$.

\subsection{Solving for amplification Factors}

To obtain the coefficients $R,I$, we solve (\ref{AppAbhinove6}). As noted earlier, this is equivalent to solving the Schrödinger like equation (\ref{AppAbhinove7}). We now consider only the slowly rotating approximation $a\omega << 1$. Defining a new coordinate $x$ given by
\begin{equation}
x = \frac{r - r_+}{r_+ - r^{\prime{}}}~, 
\end{equation} 
the radial equation can be approximately written as: 
\begin{equation}
x^2 (x+1)^2 \dv[2]{J}{x} + x(x+1)\dv{J}{x} + \left[\tilde{k}^2 x^4  - \lambda x(x+1) +Q^2\right]J = 0~. 
\label{AB1}
\end{equation}
In the above different constant are defined as
\begin{align}
Q = \frac{\omega - q\Omega_H}{4 \pi \sigma_H};~~~ 
4 \pi \sigma_H = \frac{(r_+ - r^{\prime{}})}{k(r_+) + a^2};~~~ 
\tilde{k} = \left.\f{1}{2}\pdv[2]{K}{r}\right\vert_{r_+}. 
\end{align}

Next we will follow the conventional steps which are mentioned below (see e.g. \cite{brito2020superradiance}): 
\begin{enumerate}
\item Obtain the near horizon solution of Eq. (\ref{AB1}); i.e. the solution for the regime $x<<1$. 
\item Obtain the far horizon limit of Eq. (\ref{AB1}); i.e. in the regime $x>>1$.
\item Compare the large $x$ limit of solution obtained in step 1 and small $x$ limit of solution obtained in step 2 with the demand that these solutions must be equal due to the continuity of $\psi$. This will fix the integration constants. 
\item Finally demand that the obtained solutions are asymptotically of the form given in (\ref{2Abhinove}). This will lead to the identification of the coefficients $T,I,R$. 
\end{enumerate}
For amplification factor we need only $I$ and $R$. Following the above steps these coefficients will be obtained below. 

In the near horizon regime $x << 1$ the equation (\ref{AB1}) becomes: 
\begin{equation}
x^2 (x+1)^2 \dv[2]{J}{x} + x(x+1)\dv{J}{x} + \left[Q^2 - \lambda x(x+1) \right]J = 0~. 
\end{equation}
This has a solution given by: 
\begin{equation}\label{6Abhinove}
J(x) = A_1x^{-iQ}(x+1)^{- iQ}F\left(-l,l+1,1-2iQ,-x \right)~,
\end{equation}
where $F$ is the hypergeometric function. In the above we have used the ingoing boundary condition at the horizon. Following the slowly rotating approximation, we have also used $A_{lq} = l(l+1)$. 
In the far region limit $x >> 1$ Eq. (\ref{AB1}) becomes: 
\begin{equation}
\dv[2]{J}{x} + \frac{2}{x}\dv{J}{x} + \left[\tilde{k}^2 - \frac{\lambda}{x^2}\right]J = 0~. 
\end{equation}
This has a solution of the form:
\begin{equation}\label{7Abhinove}
J(x) = C_1 e^{-i\tilde{k}x} x^{l}U\left(1-l, 2l+2,2i\tilde{k}x\right) + C_2e^{-i\tilde{k}x} x^{-l-1}U \left(-l, -2l,2i\tilde{k}x \right)
\end{equation}
where $U$ is the confluent hypergeometric function. 
Next follow step 3.
Matching the cross limits of (\ref{6Abhinove} and \ref{7Abhinove}) we find the integrations constants $C_1$ and $C_2$ in terms of $A_1$ as
\begin{equation}\label{8Abhinove}
    C_1 = A_1 \frac{\Gamma\left( 1 -2iQ\right)\Gamma\left(2l + 1 \right)}{\Gamma\left(l + 1 \right)\Gamma\left( l + 1 - 2iQ\right)}~,
\end{equation}
and
\begin{equation}\label{9Abhinove}
    C_2 = A_1 \frac{\Gamma\left( 1 -2iQ\right)\Gamma\left(-1-2l \right)}{\Gamma\left(-l-2iQ\right)\Gamma\left( -l \right)}~.
\end{equation}
Finally comparing the far solution with the asymptotic solution at infinity (given in (\ref{2Abhinove})) we have the required coefficients: 
\begin{equation}\label{10Abhinove}
    I = \frac{1}{\omega} \left[ \tilde{k}^{l+1} \frac{C_2\left(-2i\right)^{l} \Gamma \left(-2l \right)}{\Gamma \left(- l \right)} + \tilde{k}^{-l} \frac{C_1 \left(-2i\right)^{-l-1}\Gamma \left(2l + 2 \right)}{\Gamma \left(l+1 \right)} \right]~;
\end{equation}
and
\begin{equation}\label{11Abhinove}
    R = \f{1}{\o} \left[ \tilde{k}^{l+1} \frac{C_2\left(2i\right)^{l} \Gamma \left(-2l \right)}{\Gamma \left(- l \right)} + \tilde{k}^{-l} \frac{C_1 \left(2i\right)^{-l-1}\Gamma \left(2l + 2 \right)}{\Gamma \left(l+1 \right)} \right]~.
\end{equation}
Using this form for $R,I$ we calculate the amplification factor from (\ref{ampfact}). 

It must be mentioned that we encounter $\Gamma$ functions with negative arguments in the amplification factors of both the Kerr and AOS black holes. To circumvent this, we can manipulate the functions using the $\Gamma$ function reflection formula \cite{Wolfram}: 
\begin{equation}
    \f{\Gamma(s-a+1)}{\Gamma(s-b+1)} = (-1)^{b-a}\f{\Gamma(b-s)}{\Gamma(a-s)}~.
\end{equation}
for $a,b \in Z$ and complex $s$. This is a very standard technique in this context.

\end{appendix}
\end{widetext}

\bibliographystyle{apsrev}

\bibliography{bibtexfile}

\end{document}